\documentclass{aa}

\usepackage{graphicx}
\usepackage{txfonts}
\usepackage{siunitx}
\usepackage{changes}
\usepackage{packages}

\begin{document} 

\title{Triple trouble with PSR J1618$-$3921: Mass measurements and orbital dynamics of an eccentric millisecond pulsar}
\titlerunning{Mass measurements and orbital dynamics of PSR J1618$-$3921}

\author{K. Grunthal \inst{1}\thanks{\email{kgrunthal@mpifr-bonn.mpg.de}}
          \and
          V. Venkatraman Krishnan \inst{1}
          \and
          P. C. C. Freire \inst{1}
          \and 
          M. Kramer \inst{1}
          \and
          M. Bailes \inst{8,9}
          \and
          S. Buchner \inst{7}
          \and 
          M. Burgay \inst{5}
          \and
          A.~D.~Cameron \inst{8,9}
          \and
          C.-H.R.~Chen \inst{1}
          \and
          I. Cognard \inst{2,3}
          \and
          L. Guillemot \inst{2,3}
          \and
          M. E. Lower \inst{6}
          \and
          A. Possenti \inst{5}
          \and
          G. Theureau \inst{2,3,4}
          }
\institute{Max-Planck-Institut für Radioastronomie, Auf dem Hügel 69, D-53121 Bonn, Germany
\and Laboratoire de Physique et Chimie de l'Environnement et de l'Espace, Universit\'e d'Orl\'eans, CNRS, F-45071 Orl\'eans, France 
\and Nançay Radio Astronomy Observatory, Observatoire de Paris, Universit\'e PSL, CNRS, Université d’Orl\'eans, 18330 Nançay, France, 
\and Laboratoire Univers et Th\'eories, Observatoire de Paris, Universit\'e PSL, Universit\'e de Paris Cit\'e, CNRS, F-92190 Meudon, France
\and INAF-Osservatorio Astronomico di Cagliari, via della Scienza 5, I-09047 Selargius, Italy
\and Australia Telescope National Facility, CSIRO, Space and Astronomy, PO Box 76, Epping, NSW 1710, Australia
\and South African Radio Astronomy Observatory, 2 Fir Street, Black River Park, Observatory 7925, South Africa
\and Centre for Astrophysics and Supercomputing, Swinburne University of Technology, PO Box 218, VIC 3122, Australia
\and ARC Centre of Excellence for Gravitational Wave Discovery (OzGrav), Swinburne University of Technology, PO Box 218, VIC 3122, Australia
}

   \date{\today}

\abstract
   {PSR J1618$-$3921 is one of five known millisecond pulsars (MSPs) in eccentric orbits (eMPSs) located in the Galactic plane, whose formation is poorly understood. Earlier studies of these objects revealed significant discrepancies between observation and predictions from standard binary evolution scenarios of pulsar-Helium white dwarf (HeWD) binaries, especially in the case of PSR J0955-6150, for which mass measurements ruled out most eMSP formation models. }
   {We aim to measure the masses of the pulsar and its companion, as well as constraining the orbital configuration of PSR J1618$-$3921. This facilitates understanding similarities among eMSPs and could offer hints on their formation mechanism. }
   {We conducted observations with the L-band receiver of the MeerKAT radio telescope and the UWL receiver of the Parkes Murriyang radio telescope between 2019 and 2021. These data were added to archival Parkes and Nan\c{c}ay observations. We perform a full analysis on this joint dataset with a timing baseline of 23 years. We also use the data from recent observations to give a brief account of the emission properties of J1618$-$3921, including a Rotating Vector model (RVM) fit of the linear polarisation position angle of the pulsar.}
   {From the timing analysis, we measure a small but significant proper motion of the pulsar. The long timing baseline allowed for a highly significant measurement of the rate of advance of periastron of $\Dot{\omega} = \SI{0.00145(10)}{\degree\per\yr}$.
   Despite the tenfold improvement in timing precision from MeerKAT observations, we can only report a low significance detection of the orthometric Shapiro delay parameters $h_3 = 2.70^{+2.07}_{-1.47} \si{\micro \second}$ and $\varsigma =  0.68^{+0.13}_{-0.09}$. Under the assumption of the validity of General Relativity (GR), the self-consistent combination of these three parameters lead to mass estimates of the total and individual masses in the binary of $M_\mathrm{tot}= 1.42^{+0.20}_{-0.19} \si{\msun}$, $M_\mathrm{c} = 0.20^{+0.11}_{-0.03} \si{\msun}$, and $M_\mathrm{p} = 1.20^{+0.19}_{-0.20} \si{\msun}$. We detect an unexpected change in the orbital period of $\dot{P}_{\rm b} =-2.26^{+0.35}_{-0.33}\,\times\, 10^{-12}$, which is an order of magnitude larger and carries an opposite sign to what is expected from the Galactic acceleration and the Shklovskii effect, which are {\em a priori} the only non-negligible contributions expected for $\dot{P}_{\rm b}$. We also detect a significant second derivative of the spin frequency, $\ddot{f}$. The RVM fit revealed a viewing angle of $\zeta = \SI{111(1)}{\degree}$. Furthermore, we report an unexpected, abrupt change of the mean pulse profile in June 2021 with unknown origin.}
   {We propose that the anomalous $\Dot{P}_\mathrm{b}$ and $\ddot{f}$ we measure for J1618$-$3921 indicate an additional varying acceleration due to a nearby mass, i.e., the J1618$-$3921 binary system is likely part of a hierarchical triple, but with the third component much farther away than the outer component of the MSP in a triple star system, PSR~J0337+1715. This finding suggests that at least some eMSPs might have formed in triple star systems.
   Although the uncertainties are large, the binary companion mass is consistent with the $P_\mathrm{b}-M_\mathrm{WD}$ relation, which has been verified for circular HeWD binaries and also for the two HeWDs in the PSR~J0337+1715 system. Future regular observations with the MeerKAT telescope will, due to the further extension of the timing baseline, improve the measurement of $\dot{P}_{\rm b}$ and $\ddot{f}$. This will help us further understand the nature of this system, and perhaps improve our understanding of eMSPs in general.}
   \maketitle

%%%%%%%% introduction %%%%%%%%

\section{Introduction}
\label{sec:introduction}

As so-called lighthouses in the sky, pulsars are a peerless species of astronomical objects. These highly magnetised neutron stars emit a beam of an electromagnetic radiation along their magnetic poles, which is visible as a steady train of pulses at a radio telescope as the beam periodically sweeps across the observer's line-of-sight. Due to the high accuracy of atomic reference clocks and low-noise receivers in modern radio telescopes, the times-of-arrival (ToAs) of the pulses at the telescope's location are precisely recorded. The motion of the pulsar, the radio emission propagation through the interstellar medium (ISM) as well as the motion of the radio telescope through the Solar System causes the ToAs to deviate from a purely periodic behaviour. Measuring the ToAs and fitting a model to them which accounts for all these possible effects is known as pulsar timing. In particular, the timing of millisecond pulsars (MSPs), a certain sub-population of pulsars (cf.\ Sec.~\ref{sec:introduction}), allows uniquely precise measurements of the spin, astrometric and orbital parameters because these pulsars exhibit a uniquely stable rotational behaviour \citep{handbook}. 

Timing of pulsars in the Southern hemisphere experienced a step change in precision with the arrival of the MeerKAT telescope: The low system temperature ($\sim\SI{18}{\kelvin}$) of the L-band receiver, its wide spectral coverage from (856 to \SI{1712}{\mega\hertz}, thus a bandwidth of 856 MHz) and the high aperture efficiency of its $64\times\SI{13.5}{\meter}$ offset Gregorian dishes (which improve upon the Parkes Murriyang radio telescope gain by a factor of four) make MeerKAT a powerful addition to other existing radio observatories, significantly increasing the radio sensitivity in the Southern hemisphere \citep{Jonas2016}. Furthermore, the ultrawide-low band receiver (UWL) of the Murriyang radio telescope have also significantly increased its spectral coverage and sensitivity.

This work was conducted as part of the ``RelBin'' project \citep{Kramer_etal2021}, which is one of the core sub-projects of the MeerTime project, a five-year Large Survey Project \citep{Bailes_etal2020}, aiming to use the precision of the MeerKAT telescope to explore fundamental physics via pulsar timing. As outlined in \cite{Kramer_etal2021}, the main aim of ``RelBin'' is detecting or improving on the measurement of timing parameters related to relativistic effects in the orbital motion of binary systems. Due to the high precision of observations with MeerKAT, this project offers not only a wide range of tests of gravity theories \citep[e.g.][]{Hu2022}, but also improves pulsar population studies by yielding a continuously growing catalogue of precise NS mass measurements and constraining binary evolution theories \citep[e.g.][]{Serylak2022}.

The known pulsar population can be split into two large sub-groups based on their rotational behaviour and spin evolution. The so-called millisecond pulsars exhibit a rotational period of less than \SI{30}{\milli\second}, as well as a relatively low inferred magnetic field strength ($\sim 10^{8-9}\si{\gauss}$). Additionally, about 80\% of MSPs are found in binary systems, with main sequence (MS) stars, other neutron stars (NS) or white dwarfs (WD) as their companions, among the latter the Helium white dwarfs (HeWDs) are the most numerous.
 
In the current binary evolution models, these systems originate from a stellar binary, in which the more massive star already evolved into a NS. As the companion star leaves the MS and becomes a red giant, it fills its Roche Lobe and overflows it. Some of the matter in this so-called Roche-Lobe-overflow (RLO) accretes onto the NS. During this period of $\mathcal{O}(\si{\giga\yr})$, the system is detectable as a low-mass X-ray binary (LMXB). The mass transfer from the red giant to the NS also transfers orbital angular momentum to the NS, leading to a significant spin-up of the NS, such that it becomes a MSP \citep{Radhakrishnan1982, Alpar1982}. At the end of this stage, the binary consists of a MSP and a stripped stellar core, which depending on its mass evolves either into a NS or a WD \citep{BinaryStarEvolution}. 

In most observed cases, the companion is a WD; the pulsars in these systems have significantly shorter spin periods, owing to the slower evolution and longer accretion episodes associated with lighter companions. By means of detailed numerical simulations, \cite{TaurisSavonije1999} derived a relation between the binary orbital period $P_\mathrm{b}$ and the mass of a HeWD companion $M_\mathrm{WD}$ (which we will refer to as the TS99 relation). Using catalogues of known MSP-HeWD system masses and comparing them to the latest stage of simulation results, this relation has been reviewed intensely over past decades (see e.g.\ \cite{Smedley2013} \cite{Hui2018}) and usually holds for these binaries. 

The tidal interactions accompanying the RLO lead to a circularization of the binary orbit \citep{Phinney1992}, as well as to an alignment of both the pulsar's spin axis with the angular momentum axis of the orbit \citep{PhinneyKulkarni1994}. Since the companion then evolves slowly into a WD, this low-eccentricity orbit and the spin alignment should be retained at later stages.  In systems where the companion becomes a NS, the mass loss and the kick associated with the supernova event that forms the second NS will cause a significant increase in the eccentricity of the orbit ($e$), if not outright disruption, and in many cases a misalignment of the spin of the recycled pulsar with the angular momentum of the post-SN orbit \citep[for a review, see][]{Tauris2017}.

In globular clusters, interactions with passing-by external stars can disturb the circular orbits of MSP - HeWDs, which is confirmed by the large number of eccentric binary MSPs in globular clusters\footnote{For a list of pulsars in globular clusters, see \url{https://www3.mpifr-bonn.mpg.de/staff/pfreire/GCpsr.html}.}.
Apart from these cases, the majority of MSP - WD systems in the Galactic disk exhibit the expected small residual eccentricities \cite{Phinney1992}: there are no nearby stars to perturb them, and the evolution of the companion to a WD does not increase $e$. Nevertheless over the last decade, six systems with low-mass companions (which in one case are confirmed as HeWDs, \citealt{Antoniadis2016}), with $0.027 < e < 0.13$ and $22 < P_\mathrm{b} < 32$d have been discovered \citep[see Tab.~\ref{tab:emsps} and Tab.~1 in][]{Serylak2022}. These systems clearly do not follow the $e$-$P_\mathrm{b}$ relation predicted by \cite{Phinney1992} and became known as eccentric millisecond pulsar binaries (eMSPs). These systems are puzzling; their formation mechanism has not yet been fully understood \citep{Serylak2022}.

\begin{table*}[htbp]
\centering
\caption{Binary parameters for all currently known Galactic-disk eccentric MSPs.}
\begin{tabular}{cS[table-format=2.4]S[table-format=2.4]S[table-format=1.4]lllll}
	\hline \hline
	Pulsar & \text{$P$ (ms)} & \text{$P_\mathrm{b}$ (d)} & \text{$e$} & $M_\mathrm{PSR}$ (\si{\msun}) & $M_\mathrm{c}$ (\si{\msun})  & $M_\mathrm{c, theo}$ (\si{\msun})  & presumable & Refs. \\ 
    &&&&&&& nature of system & \\ \hline
    &&&&&&&& \\[-1em] 
    J1903+0327 & 2.1499 & 95.1741 & 0.4367 & 1.667(21) & 1.029(8) & - & disrupted triple & i, j \\ \hline
    &&&&&&&& \\[-1em] 
    J1618$-$3921 & 11.9873 & 22.7456 & 0.0274 & $1.20^{+0.19}_{-0.20}$ & $0.20^{+0.11}_{-0.03}$ & 0.269–0.297 & triple & e, f, here\\
    &&&&&&&& \\[-1em] 
    \hline 
    &&&&&&&& \\[-1em] 
    J1950$+$2414 & 4.3048  & 22.1914 & 0.0798 & 1.496(23) & 0.2795$^{+0.0046}_{-0.0038}$ & 0.268–0.296 & binary & a, b\\ 
    &&&&&&&& \\[-1em] 
	J2234$+$0611 & 3.5766  & 32.0014 & 0.1293 &  1.353$^{+0.014}_{-0.017}$ & 0.298$^{+0.015}_{-0.012}$ & 0.281–0.310 & binary & c, d\\ 
	&&&&&&&& \\[-1em] 
	J1946$+$3417 & 3.1701  & 27.0199 & 0.1345 & 1.827(13) & 0.2654(13) & - & binary & g, h\\
    &&&&&&&& \\[-1em] 
	J1146$-$6610 & 3.7223 & 62.7712 &  0.0074 & - & - &  0.307–0.339 & binary & k \\ 
	  J0955$-$6150 & 1.9993  & 24.5784 & 0.1175 & 1.71(2)   & 0.254(2)   & 0.271–0.300 & binary & l \\
	\hline \hline	
\end{tabular}
\tablefoot{The first columns show the pulsar period $P$ in milliseconds, their orbital period $P_\mathrm{b}$ in days and their orbital eccentricity $e$. In case mass measurements are available, the pulsar and companion mass, $M_\mathrm{PSR}$ and $M_\mathrm{c}$ respectively are given as well. For comparison we also calculate the companion mass $M_\mathrm{c, theo}$ theoretically expected from the $P_\mathrm{b}$-$M_\mathrm{WD}$ relation by \cite{TaurisSavonije1999}.}
\tablebib{Refs: (a) \cite{Knispel2015}, (b) \cite{Zhu2019}, (c) \cite{Deneva2013}, (d) \cite{Stovall2019}, (e) \cite{Edwards2001}, (f) \cite{Octau2018}, (g)\cite{Barr2013}, (h) \cite{Barr2017}, (i) \cite{Champion2008}, (j) \cite{Freire2011}, (k) \cite{Lorimer2021}, (l) \cite{Serylak2022} }
\label{tab:emsps}
\end{table*}

A possibility could be the formation in a triple system which became unstable, ejecting one of the components, as proposed for PSR~J1903+0327 \citep{Champion2008} by \cite{Freire2011} and \cite{PortegiesZwart2011}.
Intuitively, such a chaotic process should lead to a diversity of orbital configurations and companion types. However, eMSPs do not only have similar orbits, but also similar companion masses (all consistent with being HeWDs), which is seen by \cite{FreireTauris2014} and \cite{Knispel2015} as a strong indicator in favour of a deterministic process with a fixed outcome.

For this reason, five competing theories were put forward in order to explain the formation of Galactic eMSPs. They commonly rely on the TS99 relation, but describe various perturbative mechanisms capable of introducing an eccentricity of the binary orbit. A broader introduction to these can be found in \cite{Serylak2022}. Lately, the timing analysis of J0955$-$6150 \citep{Serylak2022} revealed that this system violates the TS99 relation, which is not compatible with all five theories. 

The following analysis of the eMSP PSR~J1618$-$3921 aims to broaden the knowledge about these systems, to find any similarities that could pave the way towards new formation models.

The discovery of PSR~J1618$-$3921 (henceforth J1618$-$3921, similarly all other J2000 object names refer to pulsars if not indicated otherwise) was reported by \cite{Edwards2001} as part of a 1.4-GHz survey of the intermediate Galactic latitudes with the Parkes radio telescope. It is a recycled Galactic-disk pulsar in a binary orbit with a period of \SI{22.7}{\days} and a low-mass companion, presumably accompanied by a low-mass HeWD. With a rotational period of \SI{11.98}{\milli\second}, but unmeasured period derivative, it was suspected of being an MSP. As a result of the first observations, J1618$-$3921 stood out from the pulsar population in the Galactic Plane due to its anomalously large orbital eccentricity of 0.027 \citep{Bailes2007}. It is now thought to belong to the eMSP class \citep{Bailes2007,Serylak2022}; it is however the pulsar with by far the lowest eccentricity and longest spin period within that sub-population.

After a decade of sporadic observations with Parkes, \cite{Octau2018} aimed to precisely measure the pulsar's spin, astrometric and orbital parameters via a set of dense observations of the pulsar with the Nan\c{c}ay radio telescope (NRT): 51 h of regular observations spread over three observing campaigns. This resulted in the first ever timing solution for this system, its parameters are given in Tab.~3 of \citep{Octau2018}; for completeness also shown in the second column Tab.~\ref{tab:parameters}. This shows that the pulsar is a MSP (from the small period derivative) and confirm the unusual orbital eccentricity. Due to limited precision (this means, a comparably large mean uncertainty in the Nan\c{c}ay ToAs) and timing baseline, the observations were not sufficient to reveal additional timing parameters such as the pulsar's proper motion, the rate of advance of periastron or the Shapiro delay. 

After the addition of J1618$-$3921 to the RelBin program, it has been regularly observed with the MeerKAT radio telescope. In addition to that we have also started observing it regularly with the Parkes Radio Telescope and continued observations at NRT. Using all extent data on this pulsar - adding up to a total baseline of more than 23 years - we derived an updated timing solution that improves on both numerical precision and the number of measured relativistic effects of the binary orbit, including the first estimates of the component masses.

In the course of the paper, we will start with a brief summary of the observations of J1618$-$3921 in Section \ref{sec:observations}. Section \ref{sec:analysis} will cover the profile analysis; Section \ref{sec:timing_analysis} contains the timing analyses, where we report our new timing solution, including constraints of additional parameters compared to these reported by \cite{Octau2018}, which include the constraints on the mass of the system. This will be followed in Section \ref{sec:discussion} by a thorough discussion of the current state of knowledge on eMSPs in Section \ref{sec:analysis}, with special focus on the combined results from the timing of other eMSPs and our J1618$-$3921 timing parameters. Finally, we conclude by summarising our results in Section \ref{sec:summary}.

%%%%%%%%%%% observation %%%%%%%%%

\section{Observations and data processing}
\label{sec:observations}

\begin{table}
	\centering
     \caption{Summary of all observations of J1618$-$3921 used in this work.} 
    \resizebox{\linewidth}{!}{
    \begin{tabular}{lcccc}
    	\hline \hline \\ [-1em]
    	Telescope & \multicolumn{4}{c}{Parkes} \\
    	\hline
    	Receiver & \multicolumn{3}{c}{multi-beam} & UWL \\
    	Backend & \multicolumn{2}{c}{FB 1 BIT} & CPSR-2 & Medusa  \\
    	start & Aug 1999 & Aug 1999 & Jan 2003 & Oct 2019 \\
    	finish & Sep 2001 & Oct 2001 & Jul 2005 & Apr 2022 \\
    	&&&&\\[-1em]	
    	\hline
    	$T_\mathrm{obs, tot}$ & \SI{51}{\minute} & \SI{1}{\hour}\SI{14}{\minute} & \SI{1}{\hour}\SI{49}{\minute} & \SI{28}{\hour}\SI{57}{\minute} \\
    	$f_0$ (\si{\mega\hertz}) & \num{1374} & \num{1374} & \num{1374} & \num{2368} \\
    	BW (\si{\mega\hertz}) & \num{288} &  \num{288} &  \num{288}  & \num{3328} \\
    	&&&&\\[-1em]
    	\hline
    	$N_\mathrm{chn}$ & 2 & 2 & 2 & 13 \\
    	$\bar{\sigma}_\mathrm{ToA}$ & \SI{93}{\micro\second} & \SI{23}{\micro\second} & \SI{37}{\micro\second} & \SI{32}{\micro\second} \\
    	EFAC & \multicolumn{2}{c}{0.98} & 0.28 & 0.98 \\
    	&&&  0.93 &\\
    	$\log_{10}(\mathrm{EQUAD})$ & \multicolumn{2}{c}{$-$9.77} & $-$9.17 & $-$4.29\\
    	&&& $-$4.79 & \\
    	\hline
    	&&&& \\
    	\hline \hline
    	Telescope & \multicolumn{3}{c}{Nan\c{c}ay} & MeerKAT \\
    	\hline \hline
    	Receiver &  \multicolumn{3}{c}{L-Band} & L-band  \\
    	Backend & BON &\multicolumn{2}{c}{NUPPI} & PTUSE   \\
    	start & May 2009 &  Oct 2013 & Dec 2014 & Mar 2019 \\
    	finish & Mar 2011 & Oct 2014 & Mar 2022 & Jun 2022\\	
    	&&&&\\[-1em]
    	\hline
    	$T_\mathrm{obs, tot}$ & \SI{6}{\hour}\SI{56}{\minute} & \SI{3}{\hour}\SI{26}{\minute} & \SI{48}{\hour}\SI{55}{\minute} & \SI{28}{\hour}\SI{51}{\minute}  \\
    	$f_0$ (\si{\mega\hertz}) & \num{1398} & \num{1484} & \num{1484} & \num{1284} \\
    	BW (\si{\mega\hertz}) & \num{128} & \num{512} & \num{512} & \num{776} \\
    	&&&&\\[-1em]
    	\hline
    	$N_\mathrm{chn}$ & 1 & 1 & 4 & 8 \\
    	$\bar{\sigma}_\mathrm{ToA}$ & \SI{227}{\micro\second}& \SI{193}{\micro\second} & \SI{30}{\micro\second} & \SI{6}{\micro\second} \\
    	EFAC & 0.82 & 0.50 & 0.86 &  1.09 \\
    	$\log_{10}(\mathrm{EQUAD})$ & $-$13.82 & $-$5.25 & $-$4.57 & $-$8.43 \\
    	\hline \hline
    \end{tabular}  
    }
    \tablefoot{For each campaign, the receiver and backend are listed, as well as its data span. The data set is characterised by the total observation time $T_\mathrm{obs, tot}$ and the centre frequency and bandwidth, $f_0$ and BW of the receiver. We also list the number of frequency channels $N_\mathrm{chn}$ each observed bandwidth was subdivided into, the mean ToA uncertainty $\bar{\sigma}_\mathrm{ToA}$ and the derived white noise parameters EFAC and EQUAD, which were determined with \textsc{temponest}. The CPSR-2 recorder independently records two frequency bands (\SI{1341}{\mega\hertz} and \SI{1405}{\mega\hertz}) and is thus fit with two sets of EFAC and EQUAD values, one for each band.}
	\label{tab:obs_J1618}
\end{table}

\subsection{Parkes}
The first observations of J1618$-$3921 at the Parkes Radio telescope date back to the 1999 project P309 \citep{Edwards2001}, followed by observations in 2001 during P360. In total, the pulsar was observed on six days in August 1999 and on three days in 2001, spanning the orbital phase from 0 to $0.3$ and $0.5$ to $0.7$ respectively. Both runs use the central beam of the 13-beam \SI{21}{\centi\meter} "multi-beam" receiver \citep{Staveley-Smith1996}, with a central frequency
of \SI{1374}{\mega\hertz} and a bandwidth of \SI{288}{\mega\hertz}. After a change to the CPSR-2 (Caltech-Parkes-Swinburne-Recorder) backend, J1618$-$3921 was monitored again in the first half of 2003 with a monthly cadence (covering the orbital phase between $0.2$ and $0.7$) and twice in 2005 with a gap of five days. These observations were now to made simultaneously using two different \SI{64}{\mega\hertz} bands, with central frequencies at \SI{1341}{\mega\hertz} and \SI{1405}{\mega\hertz} respectively. Further technical details of these observations are described in \cite{Edwards2001,Manchester2001}.

Making use of the ultra-wide band receiver together with the Medusa-backend \citep{Hobbs2020}, observations of J1618$-$3921 with Parkes resumed in 2019, and continue at the time of writing on a regular basis. The UWL receiver has a bandwidth of \SI{3328}{\mega\hertz} centred around a frequency of \SI{2368}{\mega\hertz}. When used in pulsar folding mode, the data have a typical sub-integration length of \SI{30}{\second} with a resolution of 128 channels per each of the 26 sub-bands, i.e.\ each channel has a bandwidth of \SI{1}{\mega\hertz}, 1024 phase bins and full polarisation information \citep{Hobbs2020}.

\subsection{Nan\c{c}ay}

As pointed out in \cite{Octau2018}, J1618$-$3921 was first observed at Nan\c{c}ay in May 2009 with the Berkeley-Orléans-Nan\c{c}ay (BON) instrument. Due to a lack of detailed information on spin, orbital parameters and the dispersion measure (DM), these first observations were conducted using the "survey" mode. The incoherent de-dispersion and coarse time resolution associated to this mode lead to very large ToA uncertainties. After the change of the Nan\c{c}ay instrumentation to the NUPPI, a clone of the Green Bank Ultimate Pulsar Processing Instrument (GUPPI) in August 2011, J1618$-$3921 was still observed in survey mode, with a total bandwidth of \SI{512}{\mega\hertz} divided in 1024 channels with a \SI{64}{\micro\second} sampling. When a coherent timing solution for J1618$-$3921 was found, observations were continued using the "timing" mode of NUPPI from December 2014 on. In this mode, NUPPI is able to coherently de-disperse the data and also samples with higher time resolution. This leads to a significant improvement in the quality of the observations, which is visible in the decrease of the mean ToA uncertainty. The observation lengths vary between 1500 and \SI{3400}{\second}, with  sub-integrations that vary between 15 and \SI{30}{\second}. In the other axes, all data files have the same resolution of 128 frequency bins, 2048 phase bins and full polarisation information.

\subsection{MeerKAT}

As part the RelBin programme \citep{Kramer_etal2021} at the MeerKAT telescope, J1618$-$3921 has been observed since March 2019, yielding a total observation time of 28.85 hours. All observations use the L-band receiver (central frequency of \SI{1284}{\mega\hertz} and an effective bandwidth of \SI{776}{\mega\hertz}) together with the PTUSE backend. All technical set-up details can be found in \cite{Bailes_etal2020}, \cite{Serylak2021} give a thorough description of the polarisation and flux calibration. The typical sub-integration length is \SI{8}{\second}, and each observation contains usually 2048 sub-integrations at a frequency resolution of 1024 channels over the full bandwidth, 1024 phase bins and the full polarisation information.

Comparing the details of the MeerKAT observations with the Parkes UWL observations, clearly the former have exceptionally low noise, resulting in outstanding quality of profile measurements. This is evident from the mean ToA uncertainty, which is almost a factor of six lower for the MeerKAT observations than for the Parkes (a full discussion of the timing procedure and ToA derivation will be given in Sec.~\ref{sec:timing_analysis}). However, the Parkes observations do reveal the structure of the pulse profile at higher frequencies. A summary of all observations is presented in Table~\ref{tab:obs_J1618}.

%%%%%%%%%%% Analysis %%%%%%%%%%%%%

\subsection{Data processing}

Following standard data reduction procedures in pulsar timing, we used the \textsc{psrchive} \citep{PSRCHIVE} software package. If not explicitly indicated otherwise, all programs or commands referred to in this section are part of this package.

The early Parkes data sets were manually cleaned from radio frequency interference (RFI) using \textsc{pazi} and \textsc{psrzap}. We used the \textsc{psrpype} pipeline\footnote{publicly available under \url{https://github.com/vivekvenkris/psrpype}} for the data reduction of the UWL observations, that have observing lengths between 2048 and 14402 seconds. \textsc{psrpype} uses the \textsc{clfd} software package\footnote{publicly available under \url{https://github.com/v-morello/clfd}} \citep{Morello2018} RFI cleaning and flux calibration measurements of the Hydra A radio galaxy, returning cleaned and flux calibrated pulsar archives. In order to polarisation calibrate the observations, METM (Measurement Equation Template Matching) \citep{vanStraten2013} was performed on the observations, using off-target calibration observations with injected pulses from a noise diode. The calibrated and cleaned UWL-data was folded into 13 frequency sub-bands.

By default, all pulsar fold-mode observations conducted with MeerKAT as part of the RelBin program are put through the \textsc{meerpipe} pipeline, which performs the RFI excision and polarisation calibration. \textsc{meerpipe} is a modified version of \textsc{coastguard} \citep{Lazarus2016}. For the polarisation calibration, a calibration observation is performed before each pulsar observation session, from which the Jones matrices used to calibrate the pulsar observations are obtained. For more details, see \cite{Kramer_etal2021}. The cleaned and calibrated files are then decimated in time, frequency and polarisation to the desired resolution, which in the case of this work means a scrunching factor of 116 in frequency, 128 in time and a full scrunch in polarisation. This leaves observations containing 8 frequency channels across the \SI{775}{\mega\hertz}.

The NRT data archives went through the full data reduction scheme described in \cite{Octau2018}. For the final analysis, we re-installed our latest ephemeris to the data and folded each observation completely in time and polarisation. These archives had a sufficient S/N to keep a resolution of four frequency channels across all observations. We used frequency-resolved templates to account for the strong profile evolution across frequency. These were generated by iteratively running \textsc{paas} on the four frequency channels. Then we obtain frequency resolved ToAs via the \textsc{pat} command.

\section{Radio emission properties}
\label{sec:analysis}

\subsection{Change of profile with frequency}

\begin{figure*}[htbp]
	\centering
	\includegraphics[width=0.32\textwidth, trim={40 50 30 60}, clip]{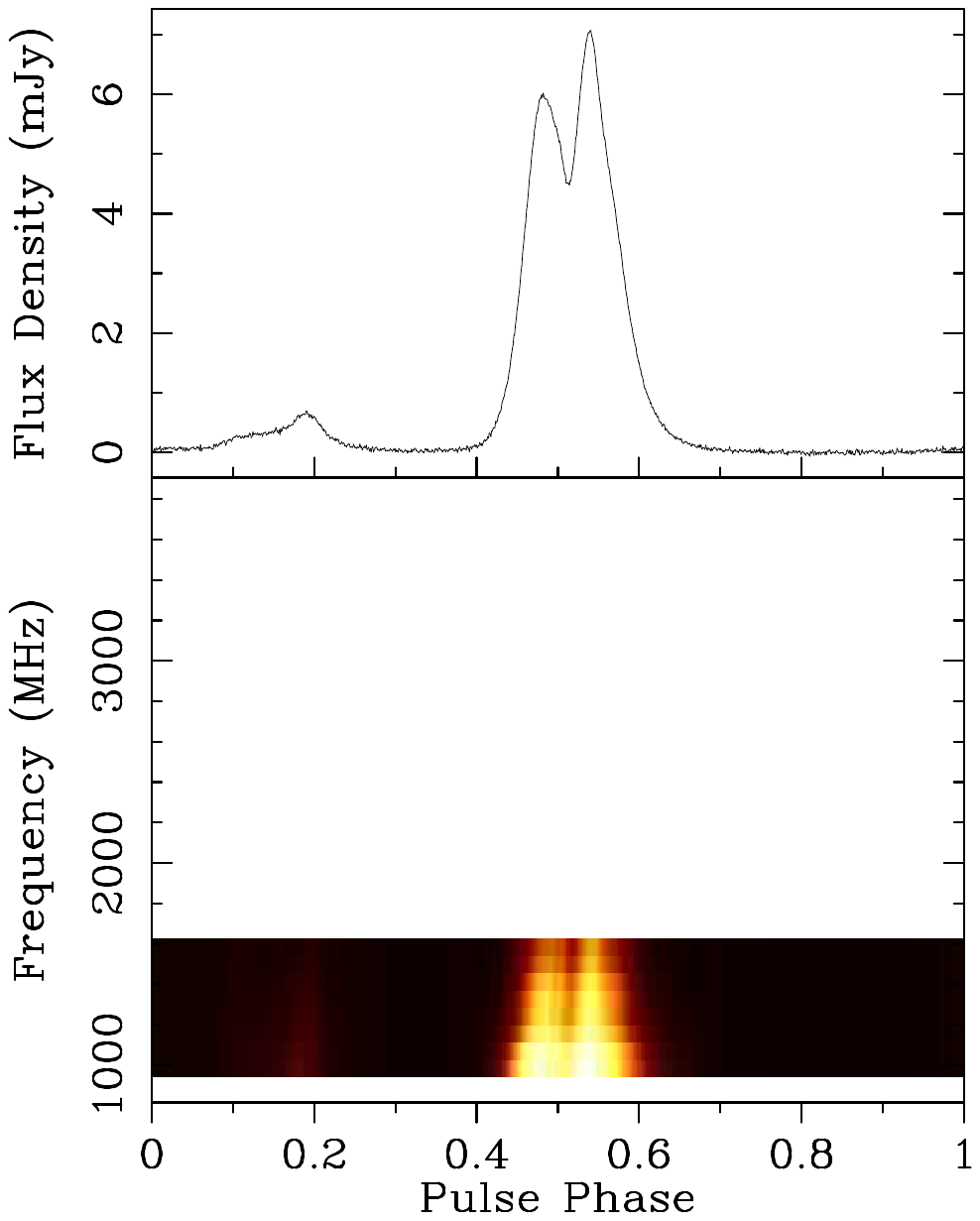}
	\includegraphics[width=0.32\textwidth, trim={40 50 30 60}, clip]{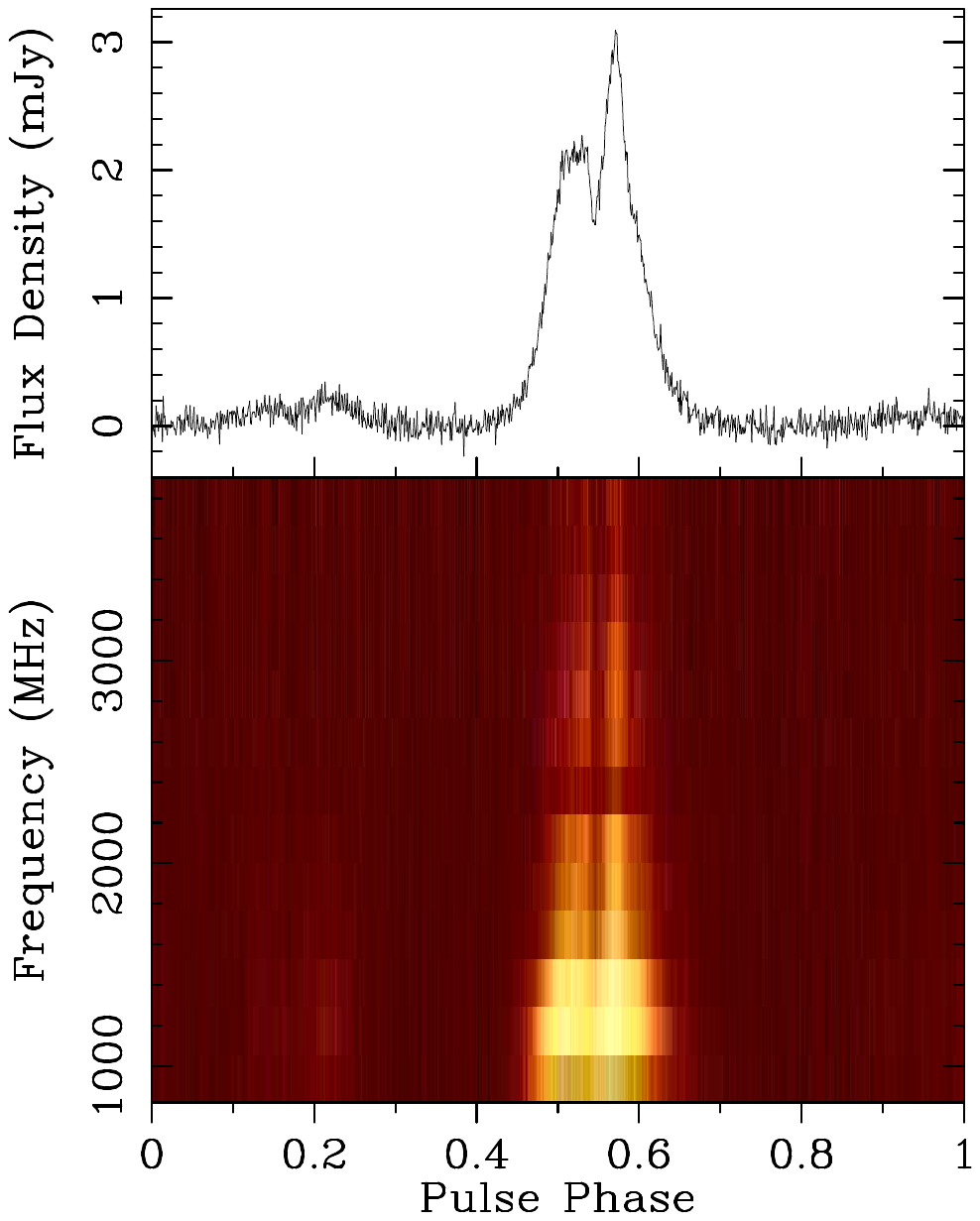}
	\includegraphics[width=0.32\textwidth, trim={40 50 30 60}, clip]{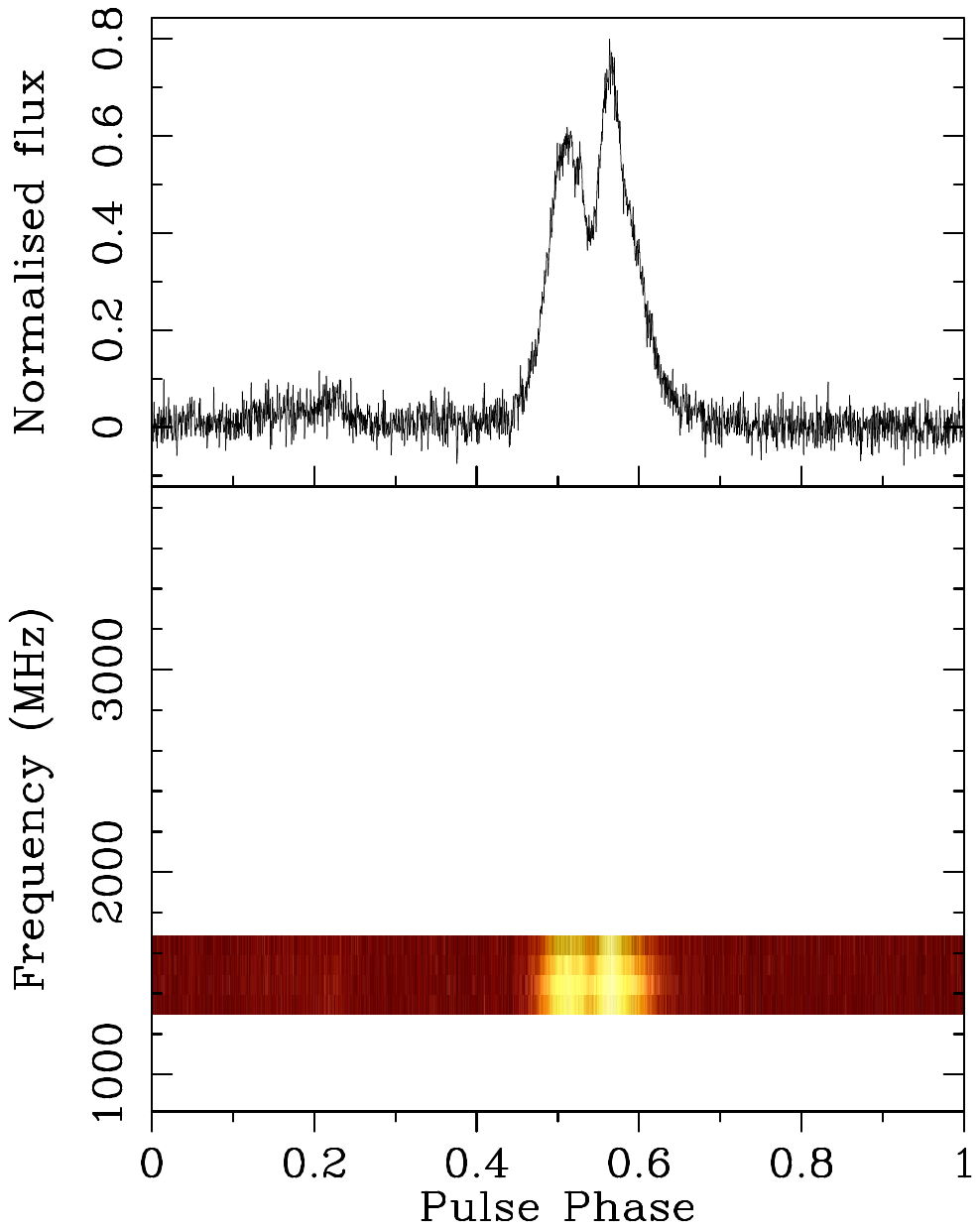}
	\caption{Frequency resolved intensity profiles from observations with the MeerKAT L-band receiver (left, $T_\mathrm{obs} \sim 28.85$ hrs), the Parkes UWL receiver (middle, $T_\mathrm{obs} \sim 28.95$ hrs) and the NRT L-band receiver (right, $T_\mathrm{obs} \sim 48.9$ hrs). The top panel of each plot shows the total intensity profile across one period, in case of the MeerKAT and Parkes observations it is flux calibrated. The NRT observations are not flux-calibrated. The bottom panels show the frequency resolved dynamic spectra across adjacent frequency bands. Note that the intensity scale of the MeerKAT and NRT dynamic spectra were adjusted to fit the range of the Parkes spectrum for ease of comparison. The MeerKAT observations were frequency scrunched to 8 channels, those from Parkes down to 13 channels and the NRT data was decimated to 4 channels. The number of channels was chosen such that the frequency resolution is kept as large as possible while providing a S/N in each band that allows for a sufficient ToA precision.}
	\label{fig:freq_intensity_plots}
\end{figure*}

If not otherwise indicated, for all analyses of the pulse profile, the integrated profile was obtained by summing up all observations of J1618$-$3921 on a backend-wise basis and summing them along the time, frequency and polarisation axes.
The left part of Fig.~\ref{fig:freq_intensity_plots} shows the profile as seen by MeerKAT's L-band receiver after $\sim 26$ hours of integration, the middle part shows the equivalent for Parkes with the UWL receiver after $\sim 29$ observing hours and the right part corresponds to the $\sim 50$ hours of observations with the Nan\c{c}ay radio telescope. The pulse profile shows a main pulse with a duty cycle of roughly 20\%. It consists of two sharper peaks, where the first one exhibits a small sub-peak on its right side. For the MeerKAT observations, the first sub-pulse peaks at $\sim6/7$ of the peak intensity of the second pulse. The main pulse is preceded by a low-intensity pulse with $1/7$ of the main pulse amplitude, which is located at $\sim \SI{110}{\degree}$ beforehand. The shape of that secondary pulse is somewhat different than that of the main pulse, with a plateau-like feature on its left-hand side and a wider peak. Although it has a duty cycle of only around 15\%, due to its low amplitude and shape plateau it appears more smeared out than the main pulse.

In all plots in Fig.~\ref{fig:freq_intensity_plots}, the heat map in the lower sub-figure resolves the pulse into the different frequency bands, a brighter colour indicating a larger intensity. Clearly, the intensity of the pulse decreases with increasing frequency, meaning that the pulsar has a steep spectrum. \cite{Spiewak2022} found a spectral index of \num{-2.28(4)}. At the same time the profile is broader at lower frequencies. For the main pulse this means that the two sharp peaks almost merge into one single broad peak at the lowest frequencies. In light of the template matching used in pulsar timing to create the ToAs, this might be a significant impairment of the ToA precision in the lower frequency bands.

\subsection{Polarisation properties}
\label{ssec:polarisation}

\begin{figure}[htbp]
    \centering
    \includegraphics[page=2, width=0.8\linewidth, trim=40 30 30 45]{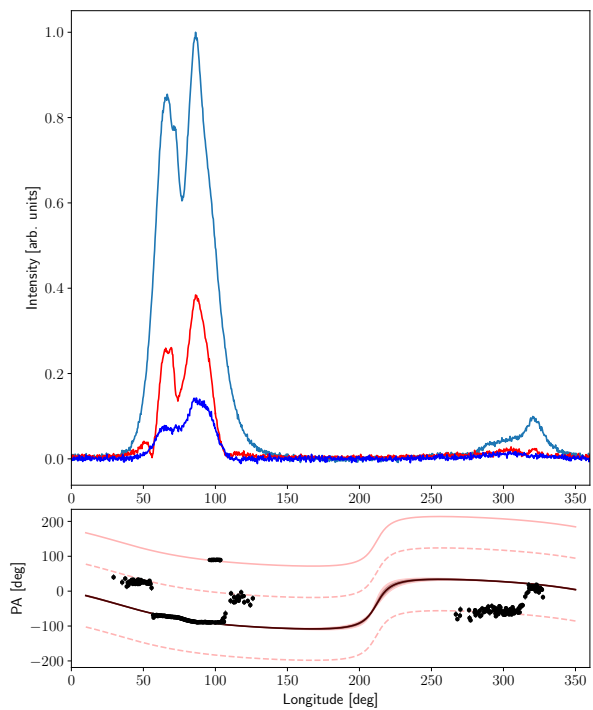}
    \caption{Polarisation profile of J1618$-$3921 obtained from integrating 29 hours of observations with the MeerKAT L-band receiver. The upper part shows the total intensity (light blue), as well as the linear (red) and circular polarisation (dark blue) fraction. The lower part shows the evolution of the position angle (PA) across the pulsar's phase. The PA exhibits the characteristic swing as well as some phase jumps. The red solid line corresponds to the Rotating Vector Model (RVM) fit to the PA, while the narrow grey band indicates the uncertainties of the fit result. The dashed line marks the RVM solution separated by 90 deg from the main one in order to include the jumped PA values. Details on the fit and the PA behaviour are discussed in Sec.~\ref{ssec:polarisation}.}
    \label{fig:mkt_pol_rvmfit}
\end{figure}

Fig.~\ref{fig:mkt_pol_rvmfit} shows the polarisation profile of J1618$-$3921 as recorded with the MeerKAT L-band receiver and corrected for the Rotation Measure given in \cite{Spiewak2022}, as well as the evolution of the position angle (PA) across the pulsar's phase. The PAs are measured in the so-called ``observer's convention''. The PA displays sudden jumps at the edges of the main pulse that are coincident with the sharp drops in the total linear polarisation. These features are consistent with arising from orthogonal polarisation modes \citep[OPMs;][]{Manchester1975a, Manchester1975b}, a phenomena that is either intrinsic to the emission of the pulsar \citep[e.g.][]{Gangadhara1997}, or result from propagation effects in the pulsar magnetosphere \citep[e.g.][]{Blandford1976, Melrose1977}.
At the right edge of the pulse we find a jump of clearly less than \SI{90}{\degree}, with an offset of only \SI{60}{\degree}--\SI{70}{\degree} from the nominal PA swing. This indicates that these jumps do not originate purely from linear modes, but most likely from magnetospheric propagation effects creating circular modes as well \citep{Petrova2001,Petrova2006, Melrose2006, Dyks2020}.

\begin{figure}
    \centering
    \includegraphics[page=1, width=\linewidth]{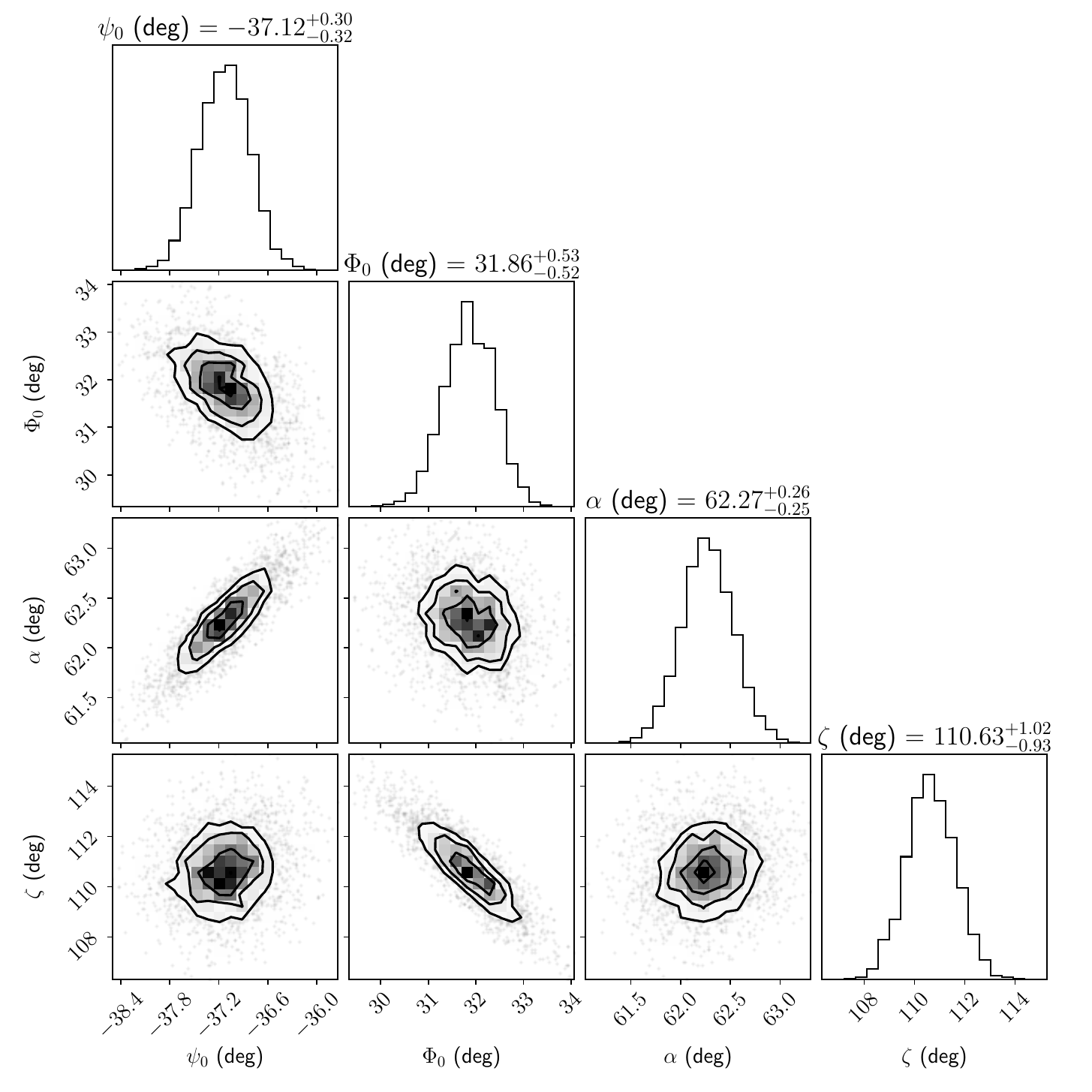}
    \caption{Corner plot showing the posterior distributions from fitting the Rotating Vector Model (RVM) to the position angle variation observed by the MeerKAT telescope.}
    \label{fig:cornerplot_rvm}
\end{figure}

We can draw information about the geometry of J1618$-$3921 from the highly resolved swing of the polarisation angle across the main pulse. This can be explained by means of the Rotating Vector Model (RVM) \citep{Radhakrishnan1969}: The emitted electromagnetic waves are polarised along the magnetic field lines, which point radially outwards along the pulsar's cone. As the beam moves across the line of sight, the observer sees these field lines under an ever changing angle \citep{handbook}.
Exploiting basic geometric considerations, the RVM yields, for the position angle $\psi$: 
\begin{equation}\label{eq:RVM}
\tan(\psi-\psi_0) = \frac{\sin\alpha\sin(\phi-\phi_0)}{\sin(\alpha+\beta)\cos\alpha - \cos(\alpha+\beta)\sin\alpha\cos(\phi-\phi_0)}
\end{equation}
where $\alpha$ is inclination angle of the magnetic axis relative to the spin axis and $\zeta$ is the angle between the line of sight and the spin axis of the pulsar. This is connected to $\beta$ (the minimum distance between the magnetic axis and the line of sight) via $\zeta = \alpha+\beta$ \citep{handbook}. This minimum distance happens at spin phase $\phi_0$; this is where $\psi$ has the steepest slope, the corresponding PA of the linear polarisation is $\psi=\psi_0$. The angles in Eq.~\ref{eq:RVM} are defined as in \cite{Radhakrishnan1969}, i.e.\ ``RVM/DT92'' convention \citep{Damour1992}. With the polarisation angle measurements from the MeerKAT observations (all data points in Fig.~\ref{fig:mkt_pol_rvmfit}), we determine the RVM parameter posteriors in their joint parameter space following the method outlined in \cite{Johnston2019}. The model also accounts for the possibility of OPM jumps and includes the corrected values in the fit. Keeping in mind the caveats associated with the RVM model, see e.g.\ \cite{Johnston2019}, the results from the best fit model are shown in terms of corner plots in Fig.~\ref{fig:cornerplot_rvm}. Following \cite{Everett2001,Johnston2019,Kramer_etal2021}, the results are presented using the RVM/DT92 convention.
We obtain $\alpha=62.27^{+0.26}_{-0.25} \si{\degree}$ and $\zeta = 110.63^{+1.02}_{-0.93} \si{\degree}$, quoting the 68\% confidence levels on the posteriors.

\subsection{Change of the profile with time}

While inspecting the timing residuals we encountered an intriguing feature in the MeerKAT observations: starting with the observation from 2021-07-06, all residuals are offset by about \SI{1}{\micro\second} with respect to all residuals before that in the data set, while the MeerKAT residuals from observations before July 2021 align with the residuals from the other telescopes after fitting for a jump between them.

\begin{figure}[htbp]
	\centering
	\includegraphics[width=\linewidth]{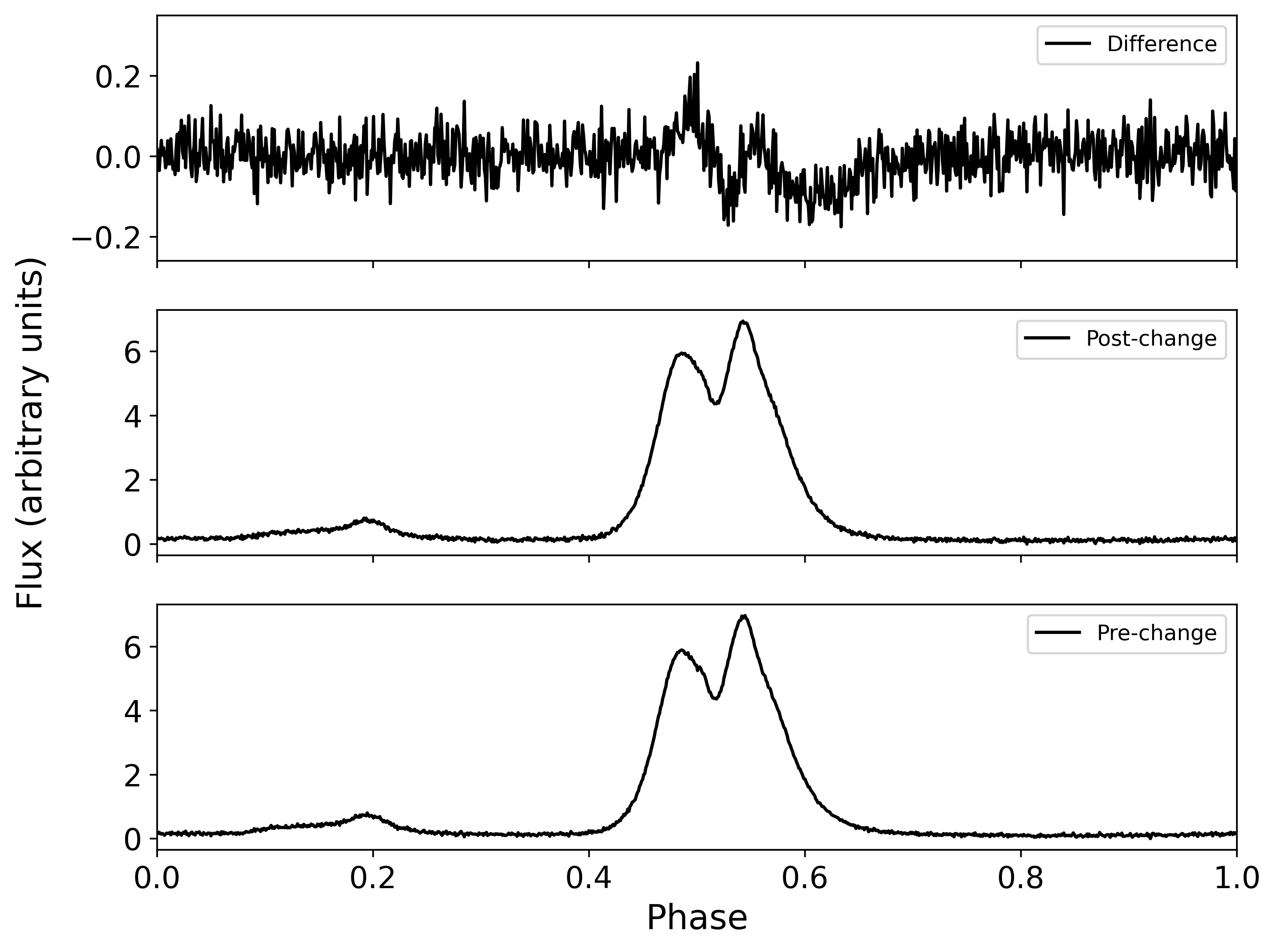}
    \caption{Phase evolution of the total profile pre- (middle panel) and post (lower panel) change. The upper panel shows the difference between both.}
	\label{fig:pre_post_pol}
\end{figure}

We found a change of the mean pulse profile to be the reason for the jump in the residuals. In Fig.~\ref{fig:pre_post_pol}, we show the summed profiles from all MeerKAT observations before the jump occurred, with a total of 26 hours, and from the 7 hours of observations since July 2022 that lead to the jumped ToAs respectively. In the following we will refer to the first one as the "pre-change profile" and to the latter one as the "post-change profile". Both profiles are generated by integrating the respective archives in time, frequency and polarisation. The first panel in Fig.~\ref{fig:pre_post_pol} contains their difference ("residual profile"), calculated by matching the pre- and post-change profile with the \texttt{ProfileShiftFit} subroutine from the \textsc{python} interface of \textsc{psrchive} \citep{PSRCHIVE} and subtracting the re-scaled version of the latter one from the first one\footnote{This numerical output and graphical display is similar to running the \texttt{pat -t -s <standard> <archive>} command.}. The underlying method of alignment is a $\chi^2$-fit of the Fourier-transformed profiles to each other to determine the respective phase shift and scale offset. The re-scaling process consists of applying the phase-rotation and overall intensity scaling of the fitting process to the latter profile. It is clearly visible, that the profiles significantly differ from each other.

To reassure ourselves that the change we see was actually occurring in June 2022, we performed a set of control analyses. To this end, we split the frequency and polarisation scrunched data from the pre- and post-change archives into two observations each. Then we repeated the subtraction procedure with these observations for all possible combinations. As expected, the fitting amongst each own data set (pre with pre and post with post) yielded flat residual profiles in both cases. When cross-correlated (pre with post and vice versa), the shape of the deviation was reproduced when correlating the profiles between the two data sets. These results indicate that we are dealing with a genuine change in the mean pulse profile from July 2021.

A few of these profile changes have been reported in the literature over the past years. One prominent example of a DM-related profile change is found in the observations of J1713+0747 \citep{Lin2021}, which was originally associated with a DM-change. A characteristic for a DM-related profile change is a $f^{-2}$ frequency dependence, i.e.\ this effect should dominate in the lower frequency bands. In contrast to that, the frequency dependence of the profile change of J1643$-$1224 \citep{Shannon2016} excluded a DM-origin. Here, \cite{Shannon2016} point to changes in the emission region of the pulsar as being accountable for a change in the emission profile. These changes in the pulsar itself are responsible for profile changes. As a DM or magnetospheric origin of the change are difficult to distinguish, we investigated the MeerKAT observations further.

\begin{figure}[htbp]
	\centering
	\includegraphics[width=\linewidth]{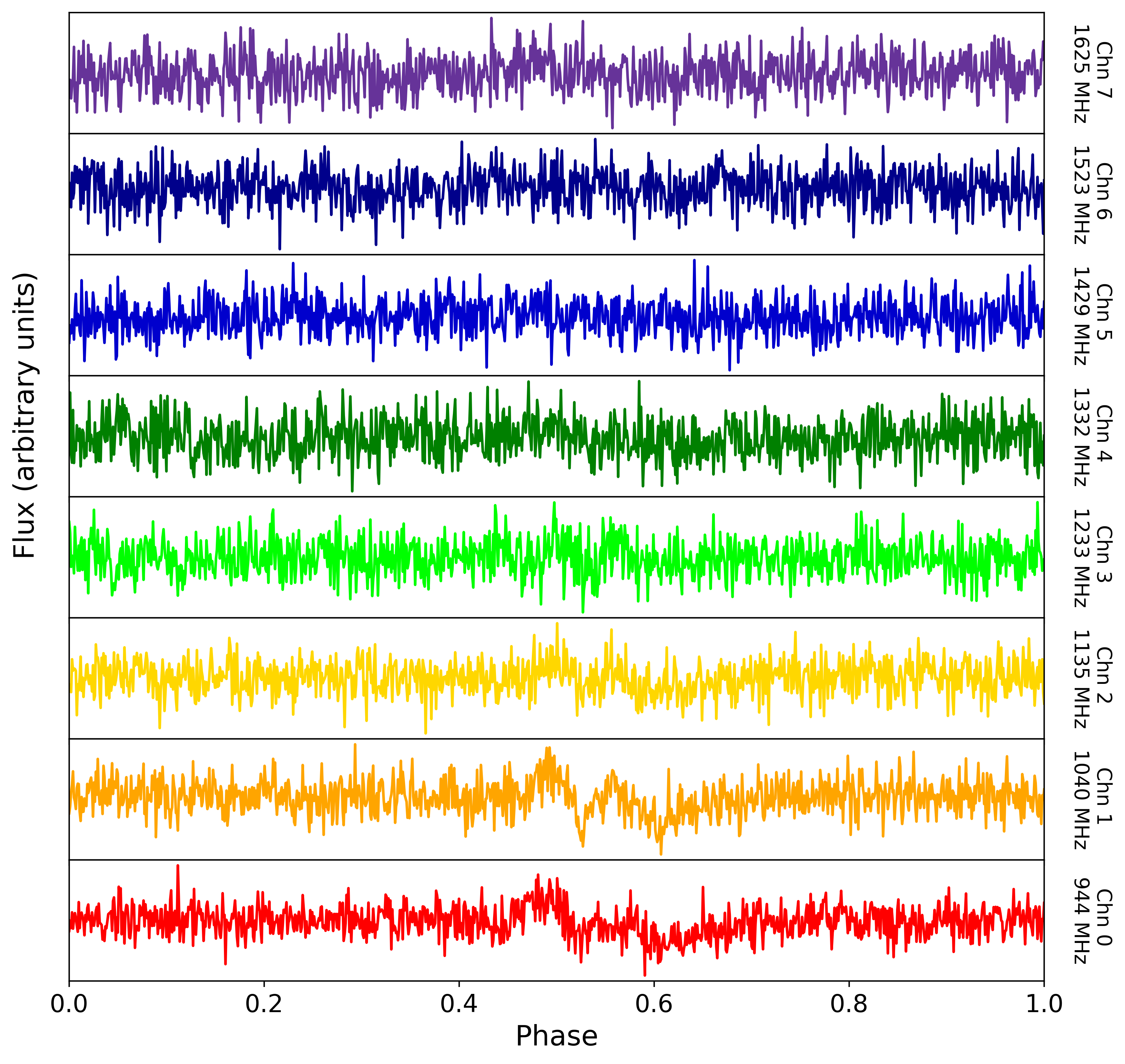}
	\caption{Difference between the "pre-change" and "post-change" profiles on a per-channel basis. The profiles were created by integrating the observations of the respective time span in time and splitting them into eight frequency sub-bands. Before subtracting, both profiles were aligned using the \texttt{ProfileShiftFit} subroutine from the \textsc{python} interface of \textsc{psrchive}.}
	\label{fig:pre_post_diff}
\end{figure}
We performed a qualitative analysis of the frequency dependence by repeating the fitting and subtraction procedure on a per-sub-band basis. In doing so, we are unfortunately limited by the S/N of the observations. As we split all MeerKAT observations into eight sub-bands, we chose to display the frequency dependence at the same resolution as in Fig.~\ref{fig:pre_post_diff}. Evidently the deviation dominates in the lower frequency bands, but the nature of the change and the available S/N prevent us to confirm or refute a $f^{-2}$ dependence.
The maximum frequency resolution feasible was sixteen sub-bands, where the deviations were most strongly visible in bands 0 to 2, weaker in bands 3 and 4, and absent from band 5 onward.

If the profile change were purely DM-related, we should be able to reproduce it by suitably altering the DM on the total pre-jump archive with the highest frequency resolution (928 channels) accordingly. After we scrunch this archive in frequency, it should give a similar residual profile as seen in Fig.~\ref{fig:pre_post_diff} when compared to the pre-jump profile with the original DM. By fitting for DM and spin frequency on a per-observation basis, we retrieve the effective change from variations in the profile. By visually inspecting the resulting DM evolution, the profile change caused an alteration of around $-\SI{0.01}{\parsec\per\cubic\centi\meter}$ in the dispersion measure. We interpret this change as not physical, but caused by the impact on the fit of the profile change. Surprisingly, a reduction of the DM in the archive header by \SI{0.01}{\parsec\per\cubic\centi\meter} in the reverse engineering scheme laid out above, did not reproduce the profile change we show in Fig.~\ref{fig:pre_post_diff}. This is a strong indicator that the profile change is caused by magnetospheric changes, rather than by the ISM.

A change in the magnetosphere or the viewing geometry might also alter the polarisation properties of the radio beam. Thus, we assessed the difference of the PA across the total profile prior to and after the jump. We did not find any indications of a change.

Putting everything together, the frequency-resolved analysis of the jump points towards a non-ISM-related profile change, as we were not able to reproduce the profile change by introducing an artificial DM change for the pre-change observations (before July 2022). We point out that we could not investigate if the change could be caused by a strong scattering event, as our spectral analysis is limited by the steep spectral index and the subsequently low S/N in the upper bands.

Since July 2021 observations of J1618$-$3921 were not only conducted at MeerKAT, but also with the Parkes and Nan\c{c}ay radio telescopes, thus we inspected the other data sets for further traces of the timing jump. With only one observation from NRT in that time span we cannot make a meaningful statement concerning any impact of the profile change. In contrast, we have several observations at the Parkes radio telescope before and after the profile change. The summed profile resulting from the Parkes observations after July 2021 does not show any significant differences to the summed profile of the observations before that date. However, the mean ToA uncertainty of the Parkes observations is much larger than the size of the respective jump needed for the MeerKAT data set. Thus we will treat these ToAs jointly.

\section{Timing analysis}
\label{sec:timing_analysis}

\subsection{Generating Times-of-Arrival}

We produced the ToAs for all data sets using the standard template matching technique employed in pulsar timing: The ToAs are calculated by correlating a standard profile against a profile the actual each observation archive over polarisation and a suitable amount of time and frequency channels. The time and frequency resolution for each telescope is chosen in a trade-off against the resulting ToA precision, resulting in the number of frequency channels specified in Tab.~\ref{tab:obs_J1618}. For frequency-resolved ToAs we created a frequency resolved standard profile by iteratively running \textsc{paas} on the integrated profile in each frequency channels. The ToAs were obtained via the \textsc{pat} command. The significant decrease of intensity in the higher frequency channels for the MeerKAT and Parkes observations result in large ToA uncertainties in these bands. For the timing analysis, we carefully discarded these ToAs in order to reduce to computational load of the analysis without altering the fit results. At most MJDs we are still left with a frequency resolution of up to 9 (7) channels for the Parkes (MeerKAT) data, which is a large improvement to the previous work \citep{Octau2018}. Due to the low S/N of the earlier data from Parkes and the NRT, those ToAs were generated using the fully integrated observations, i.e.\ one frequency channel per observation.

\subsection{Fitting timing models}

To analyse the final data set containing 1535 ToAs we use the timing software package \textsc{tempo2} \citep{tempo2}, which does a least-square minimisation of the residuals based on the $\chi^2$ statistic as well as a Bayesian noise analysis using the \textsc{temponest} plugin. In contrast to the standard \textsc{tempo2} usage, the \textsc{temponest} plugin relies on Bayesian parameter estimation, which (among other features) enables the fit for stochastic noise processes such as red timing noise and changes in dispersion measure using power law based models \citep{Lentati2014}.

The different data sets were combined by introducing a jump between each of them, with the MeerKAT data set before July 2022 as the reference data set. These jumps were treated as free fitting parameters in the \textsc{tempo2} fit, while usually being marginalised over in the \textsc{temponest} analysis. Additionally, parts of the MeerKAT data set were corrected with known jumps. 

By default, all ToA timestamps were recorded with an on-site reference clock. To be able to combine measurements from different telescopes, these are then converted to Coordinated Universal Time (UTC). Furthermore, UTC is converted to the main realisation of the terrestrial time (TT), the high-precision coordinate time standard called "International Atomic Time" (TAI, temps atomique international). It is defined via the theoretically elapsed proper time on the Earth's geoid and thus not prone to Earth's rotational variations as UTC is. Finally the ToAs are transformed to the Solar System Barycentre (SSB), by accounting for the relative motion between each telescope and the SSB with JPL's Solar System Ephemeris DE436.

For the binary orbit, \textsc{tempo2} provides several models based on the calculations by \cite{DD2} which provided a standard orbital model (henceforth "DD"). In this model, the orbital motion is parameterised by five Keplerian parameters (binary orbital period $P_\mathrm{b}$, longitude of periastron $\omega$, time of periastron passage $T_0$, orbital eccentricity $e$ and orbital semi-major axis projected along the line-of-sight $x$) and a few additional "post-Keplerian" parameters that quantify, in a theory-independent way, the relativistic deviations from the Keplerian orbital motion. Relevant here are: the rate of change of the orbital period $\dot{P}_\mathrm{b}$, the rate of periastron advance $\dot{\omega}$, the Einstein delay $\gamma$ (which quantifies the effects of the the varying gravitational redshift and special-relativistic time dilation) as well as the Shapiro delay, which affects the propagation time of the radio waves to Earth. In the DD model, the latter effect is parameterised using the "range" $r$ and "shape" parameters $s$. In GR, these are related to the companion mass $M_\mathrm{c}$ and the sine of the orbital inclination angle $\iota$ respectively \citep{Damour1992}.

Upon deriving a timing solution for J1618$-$3921 we analysed the ToAs with the theory-independent DDH model developed by \cite{FreireWex2010}, which differs from the DD model only in the parameterisation of the Shapiro delay: the new PK parameters ($h_3$ and $\varsigma$) are less correlated than $r$ and $s$, especially for systems with small orbital inclinations like J1618$-$3921.
In addition, in a later stage of the analysis, we used the "DDGR" model, which unlike the "DD" and "DDH" is not theory-independent but assumes that general relativity is the correct gravity theory, where no PK parameters are fit, only the total system mass and the companion mass. Due to the geometry of the system, the DDH model allowed for a more stable fit than the DDGR model.

After obtaining a first timing solution, which phase-connected the ToAs across the complete timing baseline, we updated the ephemeris in all available observations. With the new ephemeris installed, we repeated the entire process to obtain better profiles and standard templates. With these updated standards we then re-calculated the ToAs.

\subsection{Bayesian timing and noise models}
\label{ssec:bayesian_timing}

After deriving a final stable fit in \textsc{tempo2} with the DDH model, we performed a Bayesian non-linear fit of the timing model by means of the \textsc{temponest} software package. This plugin relies on Bayesian parameter estimation, which (among other features) enables the fit for stochastic noise processes such as white noise, red timing noise and changes in dispersion measure using power law based models \citep{Lentati2014}. Using the parameters from the \textsc{tempo2} output ephemeris as the input for \textsc{temponest}, we derived a timing solution which additionally accounted for the commonly known noise parameters: Unrecognised systematics in the ToA uncertainties are modelled by the white noise parameters EFAC $F$ and EQUAD $Q$ on a per-backend basis. Therefore the uncertainty $\sigma_\mathrm{ToA, old}$ of each ToA is re-scaled as $\sigma_\mathrm{ToA,new} = \sqrt{Q^2 + F^2\sigma_\mathrm{ToA, old}}$ \citep{Lentati2014}. For the chromatic models, we obtained an amplitude $A$ and a spectral index $\gamma$ \citep{Lentati2014}.

\begin{figure*}[htbp]
    \centering
    \includegraphics[width=\textwidth]{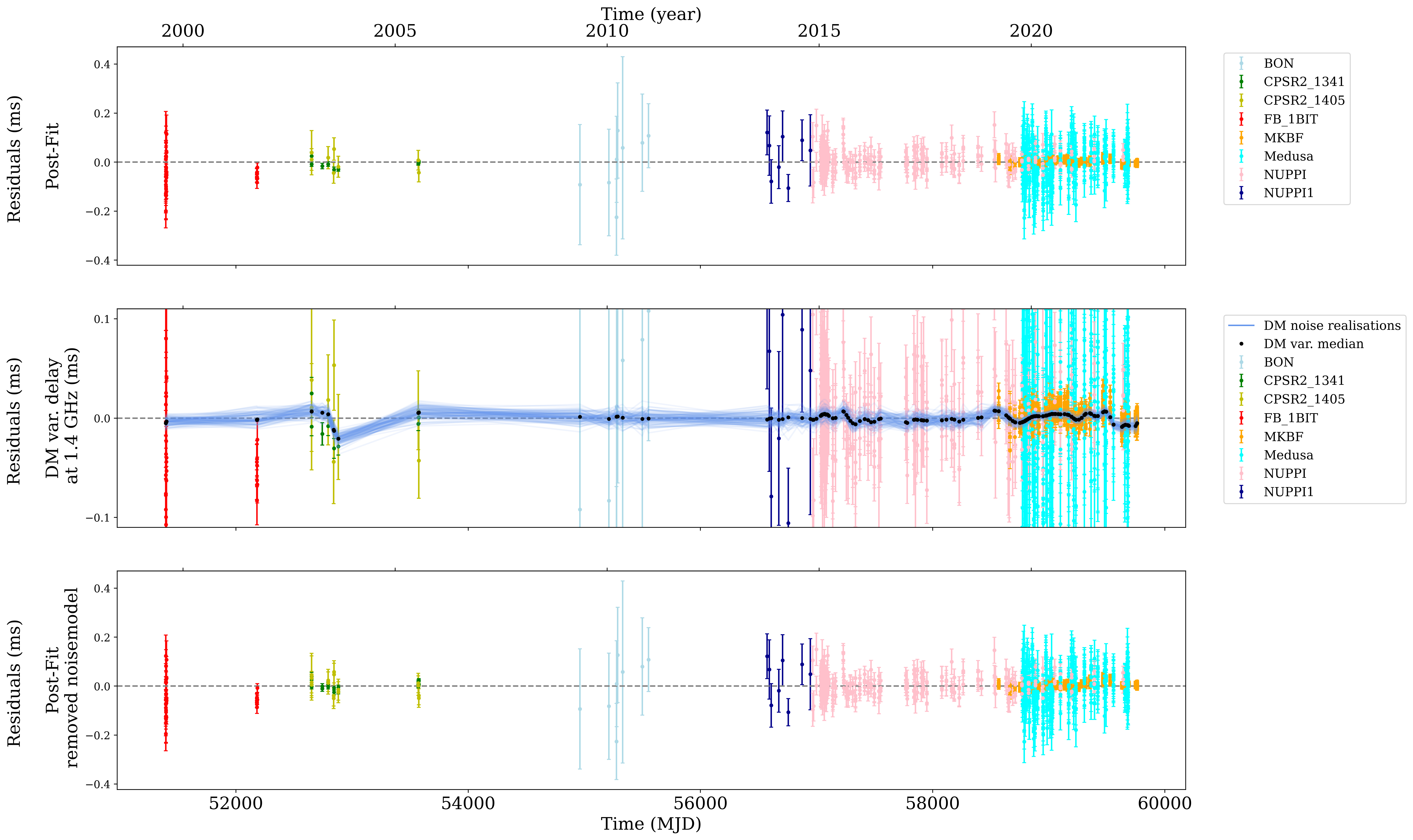}
    \caption{Upper two plots: Post-fit timing residuals as a function of time for the best-fit timing solution (\textsc{temponest} fit without removing the 100 coefficient DM noise model) of PSR J1618$-$3921 given in Tab.~\ref{tab:parameters}. Lower plot: Post-fit timing residuals after subtracting the DM noise model. \newline 
    The colours denote the different backends and systems as listed in Tab.~\ref{tab:obs_J1618}. In all plots, the uncertainties are re-scaled with the White Noise parameters EFAC and EQUAD (cf.~\ref{tab:obs_J1618}). The middle plot also contains the time domain realisation of the 100 parameter DM noise model: The blue lines show the 100 randomly created model realisations, and the black dots indicate the median across all these at each ToA.}
    \label{fig:residuals}
\end{figure*}

In order to find the best-fitting chromatic noise model, we proceeded in a two-fold way: On the one hand we compared the evidence returned by the sampler \textsc{Multinest} \citep{Feroz2019} for different combinations of noise models (Red noise (RN) only, DM noise only, Red and DM noise). On the other hand we also varied the number of noise model coefficients between 45, 60 and 100, and compared the resulting time-domain realisation between the different models. The realisations were produced using the methods of the \textsc{La Forge} github repository\footnote{Freely accessible via \url{https://github.com/nanograv/la_forge}} \citep{Hazboun2020software} adapted for the relevant models at hand. The most favoured models were the 60 and 100 coefficient DM-only models, with a difference in the log-evidences of 29. From comparing 100 averaged realisations of both noise models to the ToAs, we decided to chose the 100 coefficient model, as it visibly reflected the ToA changes more precise than the 60 coefficient model. The respective time-domain noise realisations are shown as the blue lines in the lower plot of Fig.~\ref{fig:residuals}. \textsc{temponest} accounts for the DM noise in terms of a power law model \citep{Lentati2014}, where for the chosen model we have an amplitude of $A_\mathrm{DM}=-10.37$ and a slope of $\gamma_\mathrm{DM}=0.94$. This slope is exceptionally shallow for a noise process whose slope is usually expected to be of the order of 2. From Fig.~\ref{fig:residuals} we can deduce that the residuals exhibit some significant small-timescale variations which might give rise to the shallow slope. Nevertheless, the time-domain noise realisation in the lower plot of Fig.~\ref{fig:residuals} shows that the noise model seems to match the visible trends in the data, hence we regard the noise model as satisfactory. 

As the data set exhibits large gaps in the beginning of the observations, we also investigated the covariance between the jumps and the timing parameters setting up a \textsc{temponest} analysis, where the jumps are also treated as free parameters. We did not find a significant change in any of the timing parameters.

\begin{figure*}[htbp]
    \centering
    \includegraphics[trim={30 30 30 30}, clip, width=\textwidth]{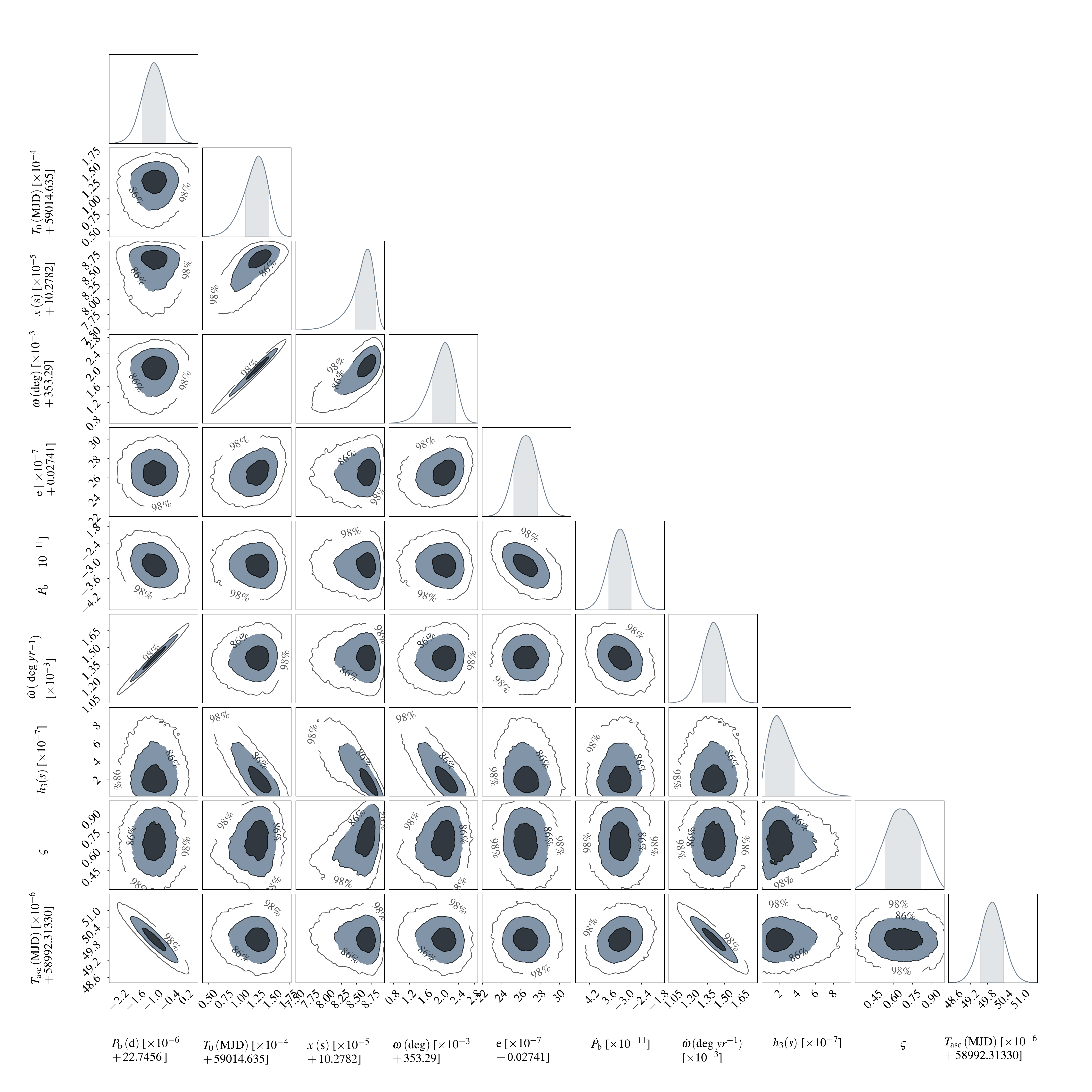}
    \caption{Corner plot for the relevant subset of timing parameters from the DDH model for J1618$-$3921 derived by applying the DDH binary models to the ToA data set applying non-linear Bayesian timing techniques using the software \textsc{temponest} to the ToAs set shown in Fig.~\ref{fig:residuals}. The diagonal elements show the 1D marginalised posterior distributions for each parameter, the shaded region indicates the 1$\sigma$ credibility interval. The 2D contours populating the off-diagonal elements show the correlation between pairs of parameters, where the lines mark the 39\%, 86\% and 98\% credibility regions, going from dark to light shaded. }
    \label{fig:TN_cornerplot}
\end{figure*}

The timing parameters of the best-fit solution from \textsc{temponest} using the DDH model are presented in the third column in Tab.~\ref{tab:parameters}. Each parameter is quoted as the maximum of the marginalised posterior together with the respective left and right $39\%$ confidence limits. The timing residuals achieved from this solution are shown in Fig~\ref{fig:residuals}. Tab.~\ref{tab:parameters} also shows the corresponding parameters reported by \cite{Octau2018}, with blank entries when the parameter was fit for for the first time in the scope of this work. In Fig.~\ref{fig:TN_cornerplot} we show both the 2D-correlation contours and the 1D posterior distributions resulting from the \textsc{temponest} analysis for a chosen subset of fitted parameters we attribute a higher relevance in this work.

In the following, we will present the individual timing parameters in greater detail and and discuss their implications for the binary system based on the numeric values derived from the best-fit \textsc{temponest} solution.

\begin{table*}[htbp]
\caption{Timing parameters from \cite{Octau2018} and the \textsc{temponest} fit performed in this work.}
\label{tab:parameters}      
\centering          
\begin{tabular}{l l l }      
\hline\hline       
Parameter & \cite{Octau2018} & This work\\ 
\hline                    
Right ascension, $\alpha$ (J2000)           & 16:18:18.8248(3)                  & 16:18:18.82500(3) \\
Declination, $\delta$ (J2000)               & $-$39:21:01.815(10)               & $-$39:21:01.832(1) \\
Reference epoch (MJD)                       & 56000                             & 59000 \\
Frequency $f$ (\si{\per\second})            &                                   & 83.421562665386(3) \\
Frequency derivative $\Dot{f}$ (\num{e-16} \si{\per\square\second})  &          & $-$3.7437(6) \\
Second frequency derivative, $\Ddot{f}$ (\num{e-27} \si{\per\cubic\second}) &   & $-$1.0(2) \\
Dispersion measure, DM (\si{\per\cubic\centi\meter\parsec}) & 117.965(11)       & 117.950$^{+0.003}_{-0.002}$\\
Dispersion measure derivative, DM1 (\si{\per\cubic\centi\meter\parsec\per\second}) &    & $-$0.0062(5) \\
Second Dispersion measure derivative, DM2 (\si{\per\cubic\centi\meter\parsec\per\square\second}) &  & $-$0.0008$^{+0.0002}_{-0.0001}$  \\
Right ascension proper motion, $\mu_\alpha$ (\si{\milli\arcsectxt\per\yr}) &    & 1.24$^{+0.14}_{-0.13}$ \\
Declination proper motion, $\mu_\delta$ (\si{\milli\arcsectxt\per\yr})   &      & $-$2.5(3) \\
Orbital period, $P_\mathrm{b}$ (\si{\day})  & 22.74559403(19)                   & 22.7455991$^{+0.0000003}_{-0.0000004}$ \\
Orbital period derivative, $\Dot{P}_\mathrm{b}$ (\num{e-11}) &                  & $-$2.26$^{+0.35}_{-0.33}$\\
Projected semi-major axis of orbit, $x$ (lt-s) & 10.278300(5)                   & 10.278285$^{+0.000001}_{-0.000002}$\\
Epoch of periastron, $T_0$ (MJD)            & 56012.21639(15)                   & 59014.635117$^{+0.000021}_{-0.000015}$ \\
Longitude of periastron, $\omega$ (\si{\degree})    & $-$6.717(3)               & 353.2919$^{+0.0002}_{-0.0003}$ \\
Longitude of periastron derivative, $\Dot{\omega}$ (\si{\degree\per\yr}) &      & 0.00142$^{+0.00008}_{-0.00010}$ \\
Orbital eccentricity, $e$                   & 0.0274133(10)                     & 0.02741231(1) \\ 
Shapiro delay amplitude, $h_3$ (\num{e-7} \si{\second})          &              & 2.70$^{+2.07}_{-1.47}$ \\
Orthometric ratio, $\varsigma$              &                                   & 0.68$^{+0.13}_{-0.09}$ \\
Span of timing data (MJD)                   & 54963.0$-$57869.1                 & 51395.2$-$55553.4\\
Number of ToAs                              & 70                                & 1535\\
Weighted residual rms (\si{\micro\second})  & 25.3                              & 8.11\\
Reduced $\chi^2$ value                      & 1.2                               & 0.91\\
\hline \hline 
Derived parameters && \\
\hline
Galactic longitude, $l$ (\si{\degree})      & 340.72                            & 340.724887 \\
Galactic latitude $b$ (\si{\degree})        & 7.89                              & 7.888043 \\
DM-derived distance (NE2001), $d$ (\si{\kilo\parsec})   & 2.7                   & 2.7 \\
DM-derived distance (YMW16), $d$ (\si{\kilo\parsec})    & 5.5                   & 5.5 \\
Rotational period, $P$ (\si{\milli\second}) & 11.987308585310(22)               & 11.98730841341(1)\\
Period derivative, $\Dot{P}$ (\num{e-20})   & 5.408(18) & 5.3796(9) \\
Total proper motion, $\mu$ (\si{\milli\arcsectxt\per\yr})  &   < 6.0            & 2.8(3) \\
Heliocentric transverse velocity, $v_\mathrm{T}$ (\si{\kilo\meter\per\second})& & 36(4) \\
Total mass, $M_\mathrm{tot}$ (\si{\msun})   &                                   & $1.42^{+0.20}_{-0.19}$ \\
Pulsar mass, $M_\mathrm{p}$ (\si{\msun})    &                                   & $1.20^{+0.19}_{-0.20}$ \\
Companion mass, $M_\mathrm{c}$ (\si{\msun})  &                                  & $0.20^{+0.11}_{-0.03}$\\
Intrinsic spin period derivative, $\dot{P}_{\rm int}$ (\num{e-20})   &          & 18(3) \\
Surface magnetic field, $B$ (\SI{e9}{\gauss}) &   0.814                         & 1.5 \\
Characteristic age, $\tau_\mathrm{c}$ (\si{\giga\yr}) &    3.5                  & 1.1 \\
Spin-down luminosity, $\Dot{E}$ (\SI{e33}{\erg\per\second}) &  1.24             & 4.1 \\
\hline\hline
\end{tabular}
\tablefoot{All uncertainties are quoted to the left and right $39\%$ confidence limits. We used the DDH model to fit for the Shapiro delay. The second column quotes the fitted and derived timing parameters from \cite{Octau2018} to the precision as given in Tab.~3 of their work. The numbers missing in the second column of the table have not been fit for by \cite{Octau2018}. The reference epoch used for position and for period differs between the previous work and this work. In the second half of the table we present quantities derived from the fit values. Opposite to \cite{Octau2018} we measure the rotational frequency and its derivatives, hence we quote to their period derivative in the second section. For the mass estimates see Sec.~\ref{ssec:mass_measurement}, for the equations to derive $B$,$\tau_\mathrm{c}$ and $\Dot{E}$ see \cite{handbook}. The last three values are derived from $\Dot{P}_\mathrm{int}$, i.e.\ they are corrected for the kinematic effects.}
\end{table*}

\begin{figure*}
	\centering
	\includegraphics[width=\textwidth]{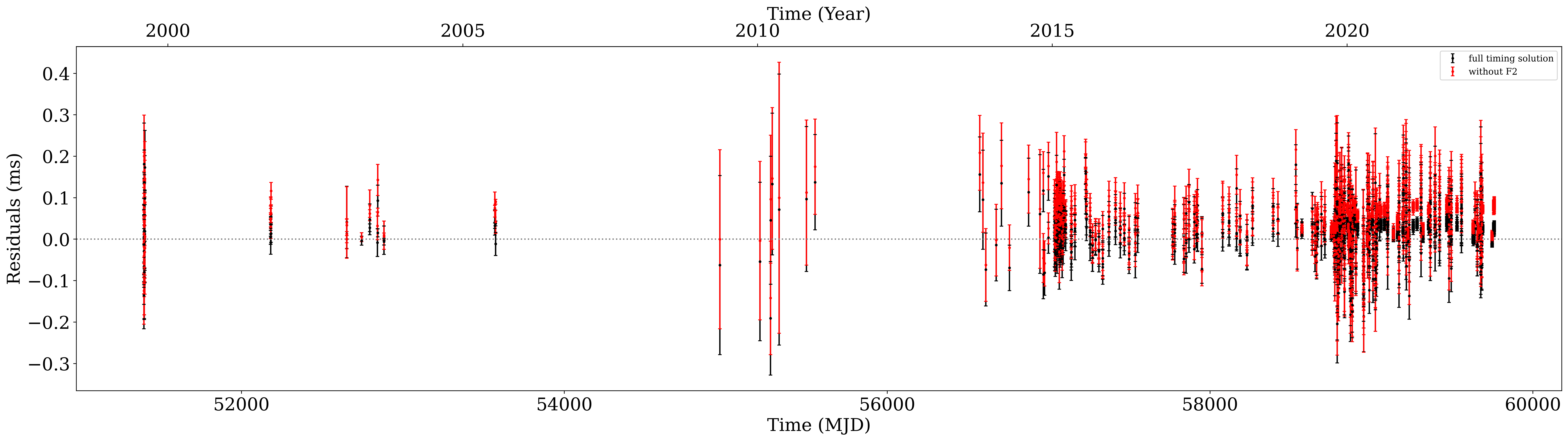}
	\includegraphics[width=\textwidth]{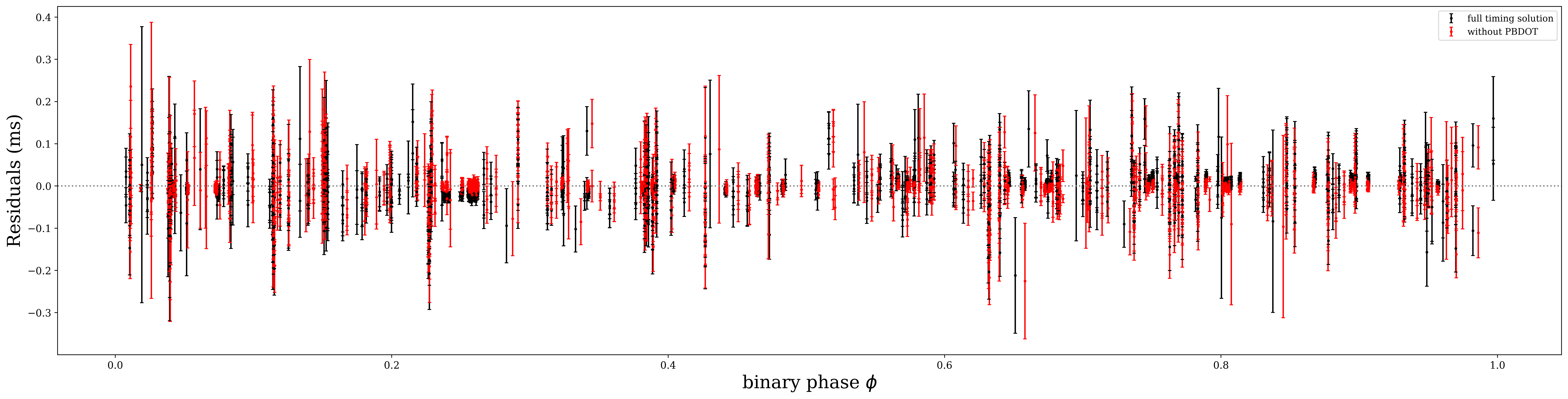}
	\caption{Top: Timing residuals as a function of time for the best-fit timing model with (black) and without (red) considering the second derivative of the rotational frequency $\Ddot{f}$. Bottom: Timing residuals as a function of orbital phase for the best-fit timing model with (black) and without (red) considering the derivative of the orbital period $\Dot{P}_\mathrm{b}$.}
	\label{fig:dropout_parameter_residuals}
\end{figure*}

\subsection{Position and proper motion}
\label{ssec:position_pm}

\begin{figure}
    \centering
    \includegraphics[width=\columnwidth]{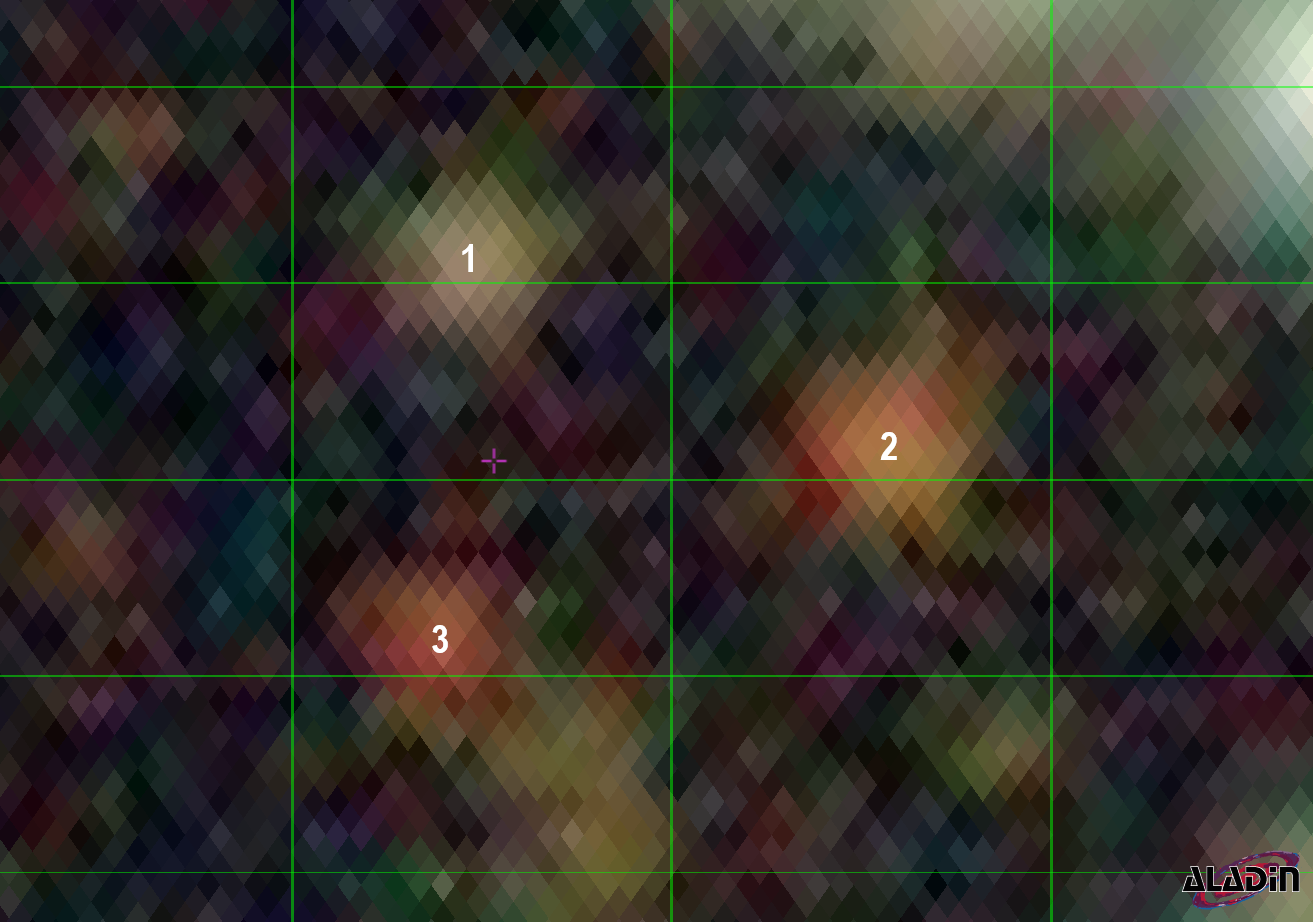}
    \caption{Excerpt from the DECaPS2 survey \citep{Schlafly2018} visualisation taken from Aladin Lite \citep{Aladinlite} in a field of view of $\sim$ \SI{17}{\arcsectxt} around the position of J1618$-$3921. The pulsar's position is indicated by the purple hair-cross. The nearest sources are enumerated with the numbers 1 to 3.}
    \label{fig:aladin}
\end{figure}

As usual, the timing solution provides the pulsar's position with very high accuracy. 
With a location at RA (J2000) 16h\,18'\,18.824940(38)'' and DEC (J2000) $-$\SI{39}{\degree}\,21'\,01.815(10)'', we searched the second data release of the DECam Plane Survey (DECaPS2) \citep{Schlafly2018}, a five-band optical and near-infrared survey of the southern Galactic plane, using the Aladin Lite web interface\footnote{\url{https://aladin.cds.unistra.fr/}} \citep{Aladinlite}. The corresponding excerpt from the survey image with a field of view of about \SI{17}{\arcsectxt} around the pulsar's position is shown in Fig.~\ref{fig:aladin}. At the position of the pulsar (indicated by the purple hair-cross on the image), we cannot identify any counterpart for either the pulsar or its companion. This implies that the electromagnetic emission of both bodies is below the detection thresholds of this survey, which are quoted to \num{23.7}, \num{22.7}, \num{22.2}, \num{21.7}, and \num{20.9} mag in the \textit{grizY} bands \citep{Schlafly2018}.

We are able to measure both the proper motion in Right Ascension $\mu_\alpha=1.24^{+0.14}_{-0.13}~\si{\milli\arcsectxt\per\yr}$ and Declination $\mu_\delta=\SI{-2.37(35)}{\milli\arcsectxt\per\yr}$. This leads to a total proper motion of \SI{-2.5(3)}{\milli\arcsectxt\per\yr}.
Furthermore, combining the timing model value of the dispersion measure DM with models of the electron distribution of the Galaxy, we infer a distance to the pulsar of 2.7 to 5.5 \si{\kilo \parsec}. For the lower boundary to the distance window we apply the NE2001 model \citep{CordesLazio2002}, the upper boundary is based on the YMW16 model \citep{Yao2017}.  Using the distance from the NE2001 model, we translate the measured proper motions into the heliocentric velocity of the binary system of $v_\mathrm{T} = \SI{33(4)}{\kilo\meter\per\second}$. 

\subsubsection{Spin-down and higher frequency derivatives}
\label{sssec:freqder}

An important quantity describing a pulsar's properties is the intrinsic spin down $\dot{P}_\mathrm{int}$. For a pulsar at a distance $d$ moving with a relative proper motion $\mu$, any time-related measurement is influenced by the change in the Doppler shift. Thus we correct the precisely measured period derivative $\dot{P} = -\SI{5.37620(68)e-20}{}$ to
\begin{equation}
\label{eq:pdot}
	\frac{\dot{P}_\mathrm{int}}{P} = \frac{\dot{P}_\mathrm{obs}}{P} + \frac{\dot{D}}{D},
\end{equation}
where $D$ is the Doppler factor caused by the unknown radial velocity of the pulsar and $\dot{D}$ its derivative. Although neither $D$ or $\dot{D}$ are known, their ratio can be estimated as: 
\begin{equation}
\label{eq:doppler_factor}
\frac{\dot{D}}{D} = - \frac{1}{c} \left[\vec{K}_0\cdot(\vec{a}_\mathrm{PSR} - \vec{a}_\mathrm{SSB}) + \frac{V_\mathrm{T}^2}{d}\right] = - \frac{a}{c} - \frac{\mu^2d}{c},
\end{equation}
where the first term holds the contribution of the line-of-sight acceleration $a$ by projecting the difference between the Galactic acceleration at the position of the pulsar $\vec{a}_\mathrm{PSR}$ and the solar system barycenter (SSB) $\vec{a}_\mathrm{SSB}$ onto the unit vector $\vec{K}_0$ pointing from the Earth to the pulsar. The second term, which depends on the transverse velocity $V_\mathrm{T}$ and the distance to the pulsar $d$, is the Shklovskii term \citep{Shklovskii1970}.

We obtain suitable values of the Galactic acceleration at the SSB and the position of the pulsar using the Milky Way mass model presented by \cite{Lazaridis2009}. For the position and velocity of the solar system barycenter we assumed $R_\odot = \SI{8.275\pm0.034}{\kilo\parsec}$ and $V_\odot = \SI{240.5\pm4.1}{\kilo\meter\per\second}$\citep{Gravity2021}.
 
Using the NE2001 distance estimate, the Shklovskii effect contributes $P\mu^2d/c=\SI{6.2e-22}{}$. The Galactic acceleration field partly compensates this effect with an excess period change of $Pa/c=\SI{-1.4e-22}{}$. We therefore arrive at an intrinsic spin-down of $\dot{P}_\mathrm{int}=\SI{5.33326e-20}{}$, which is only slightly smaller than $\dot{P}_\mathrm{obs}$ ($\dot{P}_\mathrm{int} = 0.991 \dot{P}_\mathrm{obs} $). 

Furthermore, with $\Ddot{f}= -1.0(2)\num{e-27} \si{\per\cubic\second}$ we find a non-zero value of the second derivative of the spin frequency. This value is multiple orders of magnitude larger than what is expected from a pure spin-down ($\mathcal{O}(10^{-33})$, assuming a characteristic age of \SI{10}{\giga\yr} and a braking index of 3) and among the very few values of $\Ddot{f}$ measured for the 333 pulsars with $P < \SI{30}{\milli\second}$. Outside of globular clusters, only 9 measurements of $\ddot{f}$ have been made \cite{Manchester2005}, mostly for highly energetic gamma-ray MSPs, where timing noise could be happening, additionally some of these systems are in "black widow" binaries with strong outgassing.
In one case (J1024$-$0719), the pulsar is known to have a distant companion, a K dwarf \cite{Bassa2016}, in another case, J1903+0327, the system is thought to have formed in a triple system that later became unstable \cite{Freire2011}; perhaps the third object was not fully ejected and is still somewhere in the vicinity of the system.  A comparison of the timing residuals for the timing models with and without this parameter is shown in the upper plot of Fig.~\ref{fig:dropout_parameter_residuals}. Higher derivatives of $f$ are likely to originate from a varying acceleration along the line of sight of the binary system. The implications of the measurement of $\Ddot{f}$ on the nature of the system and other timing parameters will be discussed in more detail in Sec.~\ref{ssec:pbdot_discuss}.

\subsection{Post-Keplerian parameters}

\subsubsection{Rate of advance of periastron}

The orbital eccentricity of the system and the long timing baseline allow a highly significant measurement of the rate of advance of periastron, despite the wide orbit: $\dot{\omega} =0.00142^{+0.00008}_{-0.00010}\si{\degree \per \yr}$.
If this effect is purely relativistic, it yields a direct measurement of the total mass of the system, $M_{\rm tot}$.

In order to gauge the reliability and meaning of the measurement of $\dot{\omega}$, we have to consider the possibility of additional non-relativistic effects. The most important of these is a proper motion contribution $\dot{\omega}_\mu$. This contribution is given by \citep{Kopeikin1996}
\begin{equation}
	\dot{\omega}_\mu = \frac{\mu}{\sin\iota} \cos(\Theta_\mu - \Omega),
\end{equation}
where $\Theta_\mu$ is the proper motion position angle and $\Omega$ the position angle of the line of nodes. Assuming an optimal alignment ($\cos(\Theta_\mu - \Omega)=1$), it contributes at the order of $\dot{\omega}_\mu \sim \SI{8e-7}{\degree\per\yr}$.

As discussed in \cite{Rasio1994,Joshi1997}, a third body in the system can add a contribution to the observed periastron advance: 
\begin{equation}\label{eq:omegadot_triple}
    \dot{\omega}_\mathrm{triple} = \left(\frac{\dot{x}}{x}\right)_\mathrm{triple} \frac{2\left[\sin^2{\theta_3} (5\cos^2{\Phi_3} - 1) -1\right]}{\cot{\iota}\sin{2\theta_3}\cos(\omega+\Phi_3)}.
\end{equation}
Including $\Dot{x}$ in the timing model fit yields $\Dot{x}=(2\pm8)\times10^{-15}$, which is consistent with a non-detection. Considering that the geometric terms in Eq.~\ref{eq:omegadot_triple} contribute at $\mathcal{O}(1)$, the fit value of $\Dot{x}$ gives an upper limit to the contribution of the periastron advance from the putative third body of $\dot{\omega}_\mathrm{triple} < \SI{3e-7}{\degree\per\yr}$.

Compared to the measured rate of advance of periastron, both contributions are negligibly small, so we conclude that the measured value of $\dot{\omega}$ is within measurement precision, relativistic. The relativistic $\dot{\omega}$ relates to the total mass of the system as
\begin{equation}
	M_\mathrm{tot} = \frac{1}{T_\odot} \left[\frac{\dot{\omega}}{3}(1-e^2)\right]^{3/2} \left(\frac{P_\mathrm{b}}{2\pi}\right)^{5/2},
\end{equation}
where $T_\odot\equiv {\cal GM}_\odot^\mathrm{N}/c^3= 4.9254909476412675\dots \, \rm \mu s$ is an exact quantity that follows from the exact definitions of the speed of light $c$ and the solar mass parameter ${\cal GM}_\odot^\mathrm{N}$\citep{Prsa2016}. From the best-fit parameters, we derive a total mass estimate of $1.42^{+0.20}_{-0.19}\si{\msun}$. Comparing this result with the mass measurements for similar NS-WD binaries \footnote{e.g.\ those listed under \url{https://www3.mpifr-bonn.mpg.de/staff/pfreire/NS_masses.html}}, we find that our measurement lies well within the expected mass range.

\subsubsection{Shapiro delay}

With J1618$-$3921 being a pulsar in the RelBin programme, one of the main aims of this work is achieving a significant Shapiro delay measurement by means of the high timing precision that comes along with MeerKAT observations. The rather low flux density, combined with a low inclination angle made a precise measurement of the Shapiro delay difficult. We were able to stabilise the DDH model based \textsc{tempo2} fit with ToAs gained from a dedicated superior conjunction observation campaign towards a low-significance detection of the Shapiro delay. From the \textsc{temponest} analysis we found $h_3 = 2.70^{+2.07}_{-1.47}\si{\micro \second}$ and $\varsigma = 0.68^{+0.13}_{-0.09}$. In order to convert these measurements and the measurement of $\dot{\omega}$ into constraints on the mass and the inclination angle of the system, we perform a  $\chi^2$-grid analysis of the $M_\mathrm{PSR}-\cos\iota$ space (cf.\ Sec.~\ref{ssec:mass_measurement}). The unconstrained inclination angle in the right plot of Fig.~\ref{fig:mass-mass-diagram} resulting from the analysis demonstrates that we did not arrive at a significant measurement of the Shapiro delay.

\subsubsection{Mass measurement}
\label{ssec:mass_measurement}

\begin{figure*}
    \centering
    \includegraphics[width=\textwidth, trim={10 10 50 100}, clip]{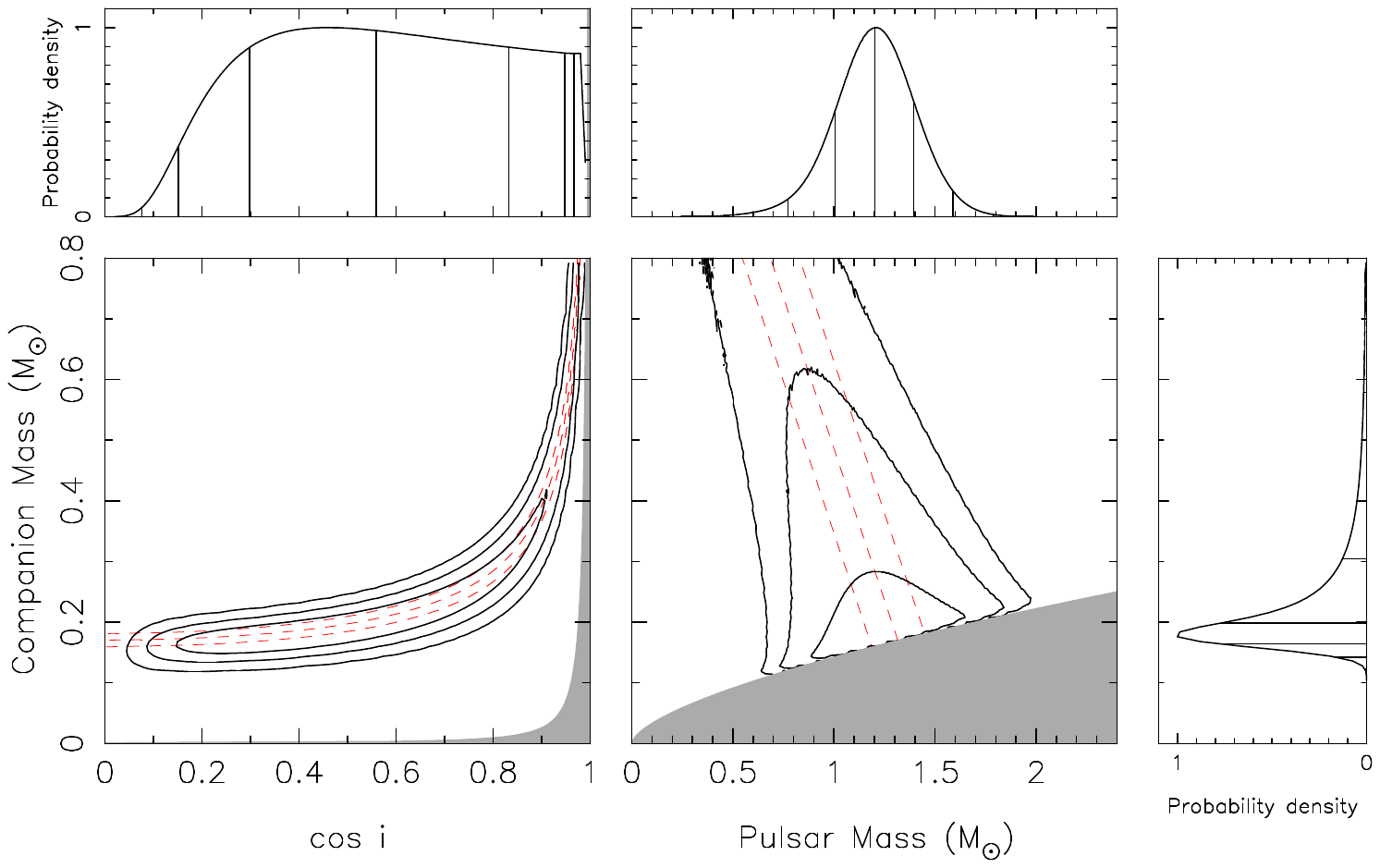}
    \caption{Constraints on the orbital inclination and masses of the J1618$-$3921 binary. The red dashed lines correspond to constraints derived from the \textsc{temponest} fit for the rate of advance of periastron $\dot{\omega}$ presented in Tab.~\ref{tab:parameters} assuming the validity of GR and regarding the effect as purely relativistic. The black lines include 68.3 and 95.4\% of all probability of the 2-dimensional probability distribution functions (pdfs) derived from the $\chi^2$ map.\\
    The left of the two inner plots shows the $M_\mathrm{c}$-$\cos\iota$ diagram, which was sampled evenly. The grey area is excluded because $M_\mathrm{PSR}$ must be positive. The plot to its right shows the projection of the $M_\mathrm{c}$-$\cos\iota$ pdf on the $M_\mathrm{PSR}$-$M_\mathrm{c}$ space using the mass function. The grey area is excluded because $\sin \iota \leq 1$. The outer three plots display the projected distributions for $\cos\iota$, $M_\mathrm{PSR}$, and $M_\mathrm{c}$. the hatched area corresponds to the $1\sigma$ intervals. The unconstrained inclination angle shows that we have a non-detection of the Shapiro delay, thus we do not provide the confidence intervals.}
    \label{fig:mass-mass-diagram}
\end{figure*}

We now estimate the masses with the highly significant detection of $\dot{\omega}$ and the weak Shapiro delay constraints using the analysis technique outlined in \cite{Barr2017}. At each grid point corresponding to a $(M_\mathrm{PSR},M_\mathrm{c})$-pair we fix the respective values of $M_\mathrm{tot}$ and $M_\mathrm{c}$ in a DDGR ephemeris adapted from the actual \textsc{temponest} results, which is then used in a \textsc{tempo2} fit. With the two mass values, the DDGR model self-consistently accounts for all observed relativistic parameters except for the orbital decay, where we know there are large contributions from other causes. The goodness of the fit is quantified by the $\chi^2$ value of the \textsc{tempo2} fit, where a lower $\chi^2$ value describes a better fit. The result is a map of $\chi^2$ values across the $M_\mathrm{PSR}$-$M_\mathrm{c}$-grid, which can be translated into credibility contours by subtracting the global minimum value across the map from all map points. The result is displayed in the mass-mass diagram in Fig.~\ref{fig:mass-mass-diagram}, together with the credibility band from the rate of advance of periastron. With this method we constrain the companion mass to $0.20^{+0.11}_{-0.03}$\si{\msun}, the pulsar mass to $1.20^{+0.19}_{-0.20} \si{\msun}$ and the total mass to $1.42^{+0.20}_{-0.19} \si{\msun}$ (68.3 \% confidence limits).

\subsubsection{Change of orbital period}

The impact of the change of the orbital period on the timing residuals is shown in the lower plot of Fig.~\ref{fig:dropout_parameter_residuals}. Similar to the measurement of the spin period derivative, the observed rate of change of the orbital period is the sum of various effects,
\begin{equation}
	\left( \frac{\dot{P}_\mathrm{b}} {P_\mathrm{b}} \right)^\mathrm{obs} =  \left( \frac{\dot{P}_\mathrm{b}} {P_\mathrm{b}} \right)^\mathrm{GW} + \left( \frac{\dot{P}_\mathrm{b}} {P_\mathrm{b}} \right)^{\dot{m}}
    -\frac{\dot{D}}{D},     \label{eq:pbdot}
\end{equation}
where apart from the kinematic contributions ($\dot{D}/D$), also emission of gravitational waves (GW) and mass loss from the system ($\dot{m}$) might significantly contribute to the measured value. Evaluating the expressions given in \cite{handbook} for the latter two effects, we find $(\dot{P}_\mathrm{b}/P_\mathrm{b})^\mathrm{GW}\sim\SI{-1e-23}{\per\second}$ and $(\dot{P}_\mathrm{b}/P_\mathrm{b})^{\dot{m}} \sim\SI{4e-28}{\per\second}$. Compared to our measured value, these contributions are negligible.

Thus, the only significant term comes from $-\dot{D}/D$. Using the value calculated in section~\ref{sssec:freqder}, we obtain $\dot{P}_\mathrm{b} = -\dot{D}/D P_\mathrm{b} \sim + 0.05 \times 10^{-12}$. Surprisingly, the best-fit timing model reveals a measured orbital period change of $-2.2^{+0.35}_{-0.33}\times10^{-11}$. This is not only two orders of magnitude larger than expected, but also carries an opposite sign. 
All considered effects are multiple orders of magnitude too small to provide an explanation for the large observed value of $\dot{P}_\mathrm{b}$. A possible solution to this tension is the presence of an additional acceleration caused by a third body in the vicinity of the binary, as discussed in Sec.~\ref{ssec:pbdot_discuss}.

\subsubsection{Other parameters}

% In contrast to
As for similar systems (cf.\ \citealt{Serylak2022}), we are not able to obtain a significant measurement of the Einstein delay amplitude $\gamma$ or any variation of the projected semi-major axis $\dot{x}$, since their contributions to the residuals are beyond the current precision of our ToAs; furthermore, given the orbital periods of these pulsars, the timing effect of $\dot{x}$ and $\gamma$ are strongly correlated \citep{Ridolfi_2019}. Moreover, we do not detect derivatives of the spin frequency higher than $\Ddot{f}$.

%%%%%%%%%%%%%% Discussion %%%%%%%%%%%%

\section{Discussion}
\label{sec:discussion}

\subsection{Comparison to Octau et al.}

In comparison to the work by \cite{Octau2018}, we use data not only from NRT, but also Parkes and MeerKAT, including observations that span back to 1999. These early observations were available previously, but only the high quality of the MeerKAT observations, together with the observation density achieved by combining three radio telescopes guaranteed a timing solution that was robust enough to extend the timing model back to 1999, through a very sparse set of observations. This large timing baseline, plus the precise recent timing, allows for the measurement of timing parameters that were not previously available: proper motion, of higher order spin and DM derivatives and post-Keplerian parameters. We are also able to significantly improve on the measurement and variation of the DM. In comparison to the four frequency channels obtained from the third NRT observation run, the large-bandwidth observations with the Parkes UWL receiver have a S/N that allows us to separate them into 13 frequency channels with often a reasonable ToA precision. Although we have to discard the ToAs from the high frequency channels, we still achieved a significant refinement in the frequency resolution compared to the previous work.

\subsection{Orbital geometry}

If the spin of the pulsar in a binary is aligned with the orbital angular momentum, the inclination angle $\iota$ coincides with the viewing angle $\zeta$. But upon comparing the timing result for the Shapiro delay parameter to the RVM fit results, there are two major caveats: First, in fitting for the Shapiro delay, we determine $\sin{\iota}$. Hence, we cannot distinguish if the corresponding inclination angle is $\iota$ or $\SI{180}{\degree}-\iota$. In case of a reliable RVM fit, this ambiguity can be solved by comparing $\iota$ to $\zeta$. This can also not be done directly, since the above RVM equation assumes that $\psi$ increases clockwise on the sky, opposite to the astronomical convention, where $\psi$ increases counter-clockwise from the above equation \citep{Damour1992, Everett2001, vanStraten2010}. Hence we have to identify the RVM fit value for $\zeta$ with $\SI{180}{\degree}-\iota$ or $\iota$ with $\SI{180}{\degree}-\zeta$ respectively \citep{Kramer_etal2021}.

Taking both these aspects into account, we first of all find with the reference angle from the RVM fit of $\SI{180}{\degree}-\zeta = 69.37^{+1.02}_{-0.93} \si{\degree}$, that $\sin{\iota}$ translates into $\iota=\SI{66(14)}{\degree}$. This is also confirmed by performing two further RVM fits in which we restricted the variation of $\zeta$ to one of the ranges allowed by the timing results (cf.\ Sec.~\ref{sec:timing_analysis}) on $\sin{\iota}$ respectively.

Although the viewing angle from the RVM fit is consistent with the inclination angle from the timing solution (Tab.~\ref{tab:parameters}), we cannot make any reasonable statement about an alignment or misalignment of both axes due to the highly unconstrained Shapiro delay.

\subsection{What causes the anomalous $\dot{P}_\mathrm{b}$ and $\ddot{f}$?}
\label{ssec:pbdot_discuss}

The significant deviation between measurement and prediction shows that there is another contribution to the pulsar's acceleration. This additional acceleration completely dominates the expected Galactic gravitational acceleration. Such a strong gravitational field could be produced by a massive nearby object. We can test this hypothesis in a simple way. If the observed $\dot{P}_\mathrm{b}$ is caused by an unexpected acceleration (and therefore implying a larger than assumed $-\dot{D}/D$ term), then we should be able to re-compute the spin-down of the pulsar using this term, as measured by $\dot{P}_\mathrm{b}$, and still obtain a positive value. Subtracting Eq.~\ref{eq:pbdot} from Eq.~\ref{eq:pdot}, and neglecting the GW emission terms, we obtain:
\begin{equation}
\dot{P}_{\rm int} = \dot{P}_{\rm obs} - P \left( \frac{\dot{P}_{\rm b}}{P_{\rm b}} \right)_{\rm obs},
\end{equation}
since $\dot{P}_{\rm b, obs}$ is negative, this has the effect of increasing our estimate of $\dot{P}_{\rm int}$ to $\sim 1.8(3) \times 10^{-19}$, which is $\sim$3.4 times larger than the observed $\dot{P}$. From this value, we estimate the characteristic age, the spin-down luminosity as well as the surface magnetic field of the pulsar \citep{handbook}. These values can be seen in Table~\ref{tab:parameters}, there we see how the change in the value of  $\dot{P}_{\rm int}$ between this work and the work from \cite{Octau2018} lead to significant differences in the values of $\tau_\mathrm{c}$ and $\Dot{E}$.

For pulsars at a low Galactic latitudes, this additional acceleration might be caused by massive molecular clouds in their vicinity. With J1618$-$3921 located at $b=\SI{7.9}{\degree}$, this is unlikely, but not impossible. Another option is that a third body is in a wide orbit around the PSR-WD binary.

The measurement of the second derivative of the spin frequency helps to distinguish between these two scenarios. A molecular cloud would be located at a large distance to the binary, thus its acceleration would appear to be constant; in this case, we would not expect large variations in the line-of-sight acceleration and thus on the $\dot{f}$. Instead, we measure a large $\Ddot{f}$ of $\SI{-1.0(2)e-27}{\per\cubic\second}$, which is very likely caused by a variation of the external acceleration. This is a strong indicator that the source of the acceleration is in the vicinity of the binary. Thus we propose that the system is a hierarchical triple system.

This line of arguments is strongly motivated by a similar discussion of the J1024$-$0719 system \citep{Bassa2016}. Upon its discovery, it was regarded as an isolated pulsar, but the measurement of higher-order spin frequency derivatives led \cite{Bassa2016} to propose a companion in an extremely wide orbit ($P_\mathrm{b} > \SI{200}{\yr}$). This was confirmed by the detection of a nearby star with the same proper motion. 
Comparing the measured value $\Ddot{f}$ for both pulsars, we find that the value for J1618$-$3921 is a factor of two smaller than for J1024$-$0719, thus of a very similar order of magnitude. With the measurement of $\Dot{P}_\mathrm{b}$ we even have the advantage of estimating the acceleration of the inner binary system - this is not possible for J1024$-$0719, because that pulsar is not already in a binary system.

Keeping in mind that most stars are part of multiple systems, it is no surprise that on rare occasions, binaries with a pulsar are actually part of a higher-order stellar system. Due to stability arguments \citep{Toonen2016}, most of these systems are hierarchical triple systems, i.e.\ they consist of an inner binary, which is in a wider orbit around a third object.

An example is the well-known triple system consisting of the MSP J0337+1715 \citep{Ransom2014}. Detailed timing of this system \citep{Ransom2014,Archibald2018,Voisin2020} revealed that both orbits of the system are co-planar and circular and the WD masses are as predicted by TS99 relation, as expected from adopting the previously discussed WD-MSP formation scenario \citep{BinaryStarEvolution}. On the other hand, \cite{Toonen2016} showed in a broad study on triple systems, that the unique dynamic in these systems also allows for a stable eccentric inner binary. They also point out, that mechanisms such as Lidov-Kozai cycles prevent a synchronisation and circularization of the binary, leading to MSP systems that stand in complete contrast to the formation scenario described by \cite{BinaryStarEvolution}.

If PSR~J1618$-$3921 really has a stellar companion, all derivatives of $f$ are expected to eventually converge on a Keplerian orbit for the outer component \citep{Rasio1994}. Here, J1024$-$0719 again serves as a precedent; we should consider these MSP companions to be also settled in exceptionally wide orbits. Any associated parameter derivative is therefore expected to show up only in data sets with a combination of a long timing baseline and significant timing precision. Determining the orbital configuration of the outer companion would require the knowledge of at least the first five derivatives of $f$ \citep{Rasio1994,Joshi1997}\footnote{The first derivative of $f$ generally cannot be used as intended by these authors, because of the {\em a priori} unknowable pulsar spin-down, but also because, in the system studied in these works (PSR B1620$-$26), the acceleration caused by the host globular cluster (M4) is also hard to estimate, given the lack of a precise 3-D position of the pulsar relative to M4. However, as mentioned before, in the case of PSR~J1618$-$3921, we have direct access to the acceleration of the system via $\dot{P}_\mathrm{b}$, which means that the equations of \cite{Joshi1997} can indeed be used.}. With the knowledge of fewer derivatives, we can only put a few constraints on the orbit \citep{Bassa2016}: $\Dot{P}_\mathrm{b}$ relates to the corresponding acceleration from the third body $a$ as $a/c\sim\Dot{P}_\mathrm{b}/P_\mathrm{b}$. Similarly $\Ddot{f}$ relates to the change of the acceleration as $\Dot{a}/c\sim \Ddot{f}/f$. From the acceleration and its change, we can place an order-of-magnitude estimate on the orbital period of the third body as $P_{\mathrm{b},3}\sim a/\Dot{a} \sim \SI{300}{\yr}$, given the values form our best-fitting timing solution. This is not unexpected, and also highly in line with the findings from \cite{Bassa2016} in the case of J1024$-$0719.

\subsubsection{Optical counterpart}

We consulted the DECaPS2 \citep{Schlafly2018} catalogue to search for a spatially resolved object that could be associated with the PSR~J1618$-$3921 system, and thus be identified as the binary companion or the putative third body. As mentioned in  Section~\ref{ssec:position_pm}, no counterparts are identified near the position of PSR J1618$-$3921.
The  upper mass limit of any companions (either the binary companion to the pulsar, or the more distant object) can thus be estimated with the depth of the catalogue through comparisons with the expected colours and magnitudes from stellar evolutionary models.
We have used the PAdova TRieste Stellar Evolutionary Code \citep[PARSEC v2.0][]{Bressan2012,Nguyen2022} to obtain the $grizY$ magnitudes in the ABmag system to facilitate comparisons with the DECaPS2 catalogue.
Applying an extinction $A_V\sim0.2$ mag \footnote{estimated via Galactic Dust Reddening and Extinction {\url{https://irsa.ipac.caltech.edu/applications/DUST/}}.} and adopting a distance of 5.5 kpc, a 0.56$\rm M_\odot$ dwarf star ($\sim$M0V,\citealt{PM13})
would have $grizY=$ 23.9, 22.5, 21.7, 21.3, 21.2 mag, respectively.
Such a star would be near the detectability limit of the \textit{riz} bands in the DECaPS2 survey, given its limiting magnitudes \num{23.7}, \num{22.7}, \num{22.2}, \num{21.7}, and \num{20.9} mag in \textit{grizY} bands respectively, and would have been detected in all 5 bands if a smaller distance of 2.7 kpc is adopted. To summarise, any companion at the location of PSR~J1618$-$3921 would have a limiting magnitude  detection threshold of \SI{23.5}{\mag}  in the $G$-band, which at the distance to the pulsar of \SI{5.5}{\kilo\parsec} corresponds to an absolute magnitude  $>\SI{9.79}{\mag}$. This could be a M-dwarf of mass $< \, 0.56\, \rm M_\odot$ or a compact object.

\subsubsection{Nearby stars, their motions and their gravitational accelerations}

We consulted the \textit{Gaia} DR3 \citep{GAIA_DR3}\footnote{\url{https://gea.esac.esa.int/archive/}} to search for objects that might have a proper motion similar to that of PSR~J1618$-$3921, i.e., within the $\pm 3\sigma$ error ellipse.
This was, incidentally, how the distant binary companion of PSR~J1024$-$0719 (a K7V star) was identified \cite{Bassa2016}.
No objects with such a proper motion are detected within a radius of \SI{1.4}{\arcminute} around PSR~J1618$-$3921. Using the NE2001 distance for a lower limit, this corresponds to a minimum distance of \SI{0.8}{\parsec}.

In the deeper DECaPS2 catalogue \citep{Schlafly2018} we find three nearby stars; shown in Fig.~\ref{fig:aladin}, at a distance of 2´´, 4´´ and 2´´ (following the labels 1 to 3) from PSR~J1618$-$3921. Given the depth of this catalogue, these faints stars are not in the \textit{Gaia} DR3 catalogue, so an association with PSR~J1618$-$3921 cannot be excluded based on proper motion measurements.
Under the assumption that the three objects are stellar type objects and that they are at the same distance as the pulsar, we have extracted their \textit{grizY} magnitudes from the DECaPS2 catalogue to estimate their masses.
We use Star 1 as an example as it has measurements in all 5 bands: 23.3, 22.0, 21.3, 20.7, 20.5 mag respectively.
These magnitudes are in agreement with those for a \SI{0.6}{\msun} (or K9V) star with $A_V\sim0.2$ and a distance of 5.5 kpc: 23.4, 22.0, 21.3, 21.0, and 20.9 mag respectively.

\begin{figure}
    \centering
    \includegraphics[width=\linewidth]{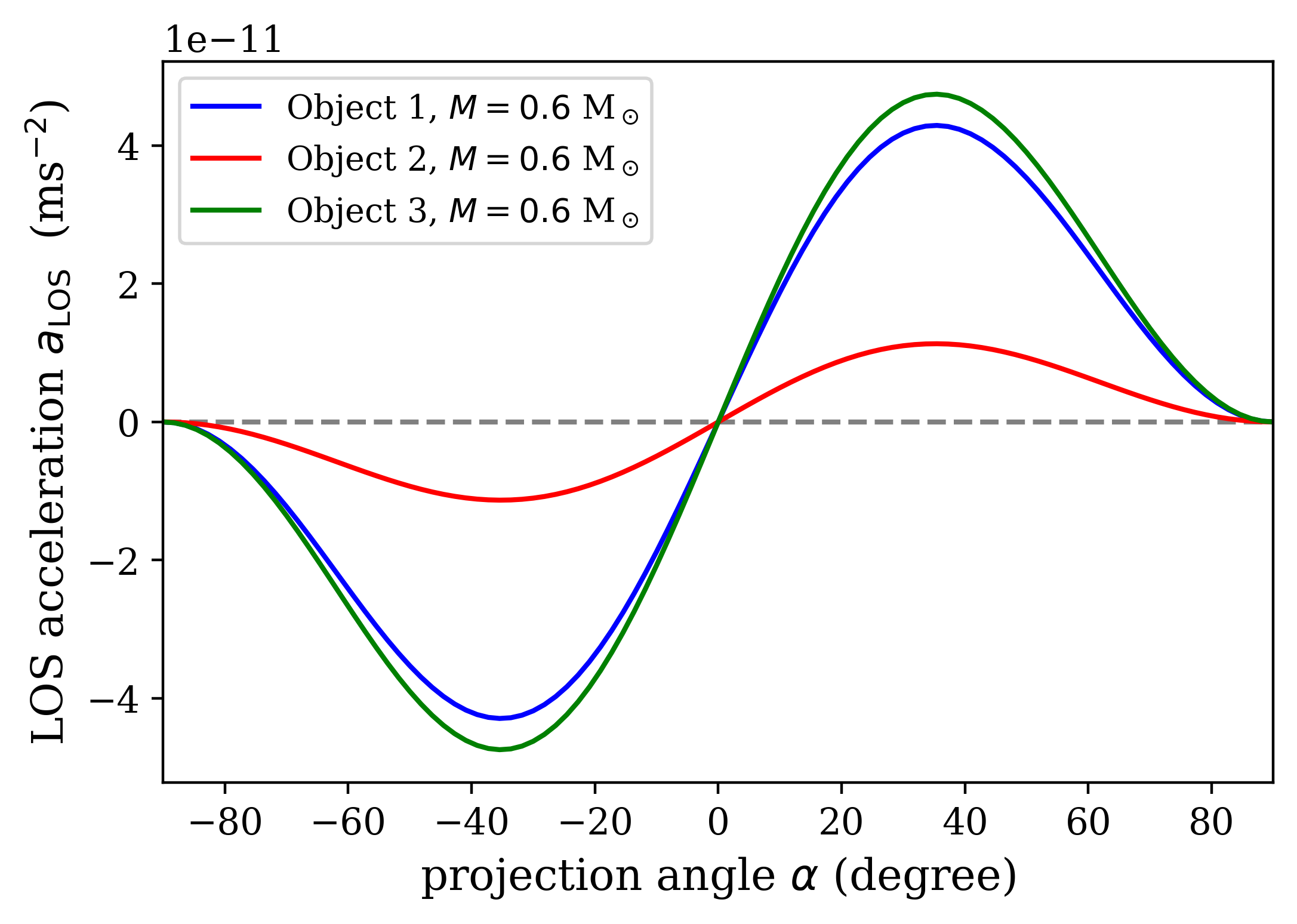}
    \caption{LOS acceleration $a_\mathrm{LOS}$ as a function of the projection angle $\alpha$ onto the LOS for the three objects in the DECaPS2 \citep{Schlafly2018} catalogue found closest to the position of J1618$-$3921. The acceleration was calculated from Newton's first law assuming they are K- or M-stars with a mass of \SI{0.6}{\msun}.
    }
    \label{fig:LOSacc}
\end{figure}

For each of the three stars, we make an order-of-magnitude estimate of the line-of-sight (LOS) acceleration $a_\mathrm{LOS}$ they exert on the pulsar respectively. Assuming a typical mass of \SI{0.6}{\msun} for these stars, we derive the estimate via Newton's law $a_\mathrm{LOS} = \frac{GM}{d^2} \sin{\alpha}$, where $M$ denotes the mass of the star, $G$ is the gravitational constant, $d$ is the distance between the pulsar and the star and $\alpha$ is the angle between the vector pointing from the star to the nearest point on the LoS and the vector pointing from the star to the pulsar. Turning the angular distance taken from \textsc{Aladin} into the physical separation, we use the NE2001 distance, as it gives us an upper limit on the acceleration. This inferred separation is the projected distance $r$ between the pulsar and the star, so $d=r/\cos{\alpha}$. The resulting acceleration curves calculated under the previously outlined assumptions for the three objects marked in Fig.~\ref{fig:aladin} are shown in Fig.~\ref{fig:LOSacc}.

All these objects cause accelerations which are roughly two orders of magnitude smaller than the acceleration $a_{\mathrm{LoS, }P_\mathrm{b}}=\SI{-3.45e-09}{\meter\per\square\second}$ obtained from $\Dot{P}_\mathrm{b}$. 
Hence the putative wide-orbit companion of J1618$-$3921 must be closer than these objects, and must have a luminosity below the DECaPS2 \citep{Schlafly2018} limit: as mentioned above, it could be an M-dwarf or a compact object.

%%%%%%%%%%% Summary %%%%%%%%%%%%%%%

\section{Summary}
\label{sec:summary}

This paper presents a comprehensive overview of the latest knowledge about the eccentric millisecond pulsar J1618$-$3921 using the combined data set from 23 years of observations with Parkes, NRT and MeerKAT radio telescopes and their respective different back-ends.

We present a detailed study on the pulsar's emission properties with two notable results: First we recorded a profile change that happened around June 2021 with the MeerKAT observations. Our analyses favours an intrinsic profile change over an ISM-related influence, but due to the limited S/N in the upper MeerKAT frequency bands, we cannot finally determine the origin of this change. Furthermore we analysed the behaviour of the position angle of the linear polarisation. Assuming a purely dipolar radio emission, with the PA perfectly following the RVM, we constrained the position of the spin axis of the pulsar to \SI{111(1)}{\degree}. The uncertainty in the orbital inclination precludes any conclusions on the alignment of the spin axis of the pulsar with the orbital angular momentum.

While in previous publications \citep{Bailes2007,Octau2018}, orbital and then phase-coherent timing solutions were already published, here we not only report the old timing with significantly improved precision, but we provide the first solution including a binary model with an increased number of Post-Keplerian parameters. The stability of the solution is mainly provided by the dense accumulation of data points from joint MeerKAT, Parkes and NRT observations in the recent past. This allowed us to include all available observations up to the very first observations from 1999. This large timing baseline significantly improved the measurement of rate of advance of periastron.

Although the ToAs obtained from monthly observations with the MeerKAT L-band receiver exhibit an outstanding precision compared to ToAs resulting from concurrent observations at the Parkes and Nan\c{c}ay radio telescopes, the low S/N of the pulsar as well as the shallow inclination angle impeded a high-significance detection of the Shapiro delay. Nevertheless we are able to present a first constraint on the orthometric parameters $h_3$ and $\varsigma$. Combining the low-significance Shapiro delay detection with the precise measurement of the rate of advance of periastron we are able to present the first ever mass estimates of this system. Unfortunately, the steep spectral index prevents us from obtaining more precise ToAs using the S-band (1.75 to \SI{3.5}{\giga\hertz}) receiver at MeerKAT. With a factor of two to three improvement in timing precision, the Shapiro delay should  be measurable with useful precision, but this will only be possible with future radio telescopes of even higher sensitivity.

The most remarkable result of the timing analysis is the amount of change of the orbital period and the large second derivative of the spin frequency, which indicate that the pulsar is actually part of a triple system.
The possibility of the evolution of J1618-3921 as a triple system opens the door for similar evolution of other eMSPs. However, there is at the moment no clear evidence that other eMSPs are part of hierarchical triple systems.

Our long-term plan for this pulsar consists of regular observations of J1618$-$3921 with the L-band receiver at MeerKAT and the UWL receiver at the Murriyang Parkes radio telescope. We expect that the increased timing baseline will significantly improve all currently measured parameters, but also enable the detection of additional parameters such as $\Dot{x}$ (which will constrain the orbital orientation of the system) or higher derivatives of $f$, which will provide additional information on the companion mass and its orbit.

%%%%%%%%%%%%%%% Acknowledgements %%%%%%%%%%%%%%%

\begin{acknowledgements}
The authors thank Aurélien Chalumeau, Michael Keith and Aditya Parthasarathy for the fruitful discussions and support regarding the noise model comparison and time domain realisation. All authors affiliated with the Max-Planck-Gesellschaft (MPG) acknowledge its constant support. VVK acknowledges financial support provided under the European Union's Horizon Europe 2022 Starting Grant ``COMPACT" (Grant agreement number: 101078094, PI: Vivek Venkatraman Krishnan). The MeerKAT telescope is operated by the South African Radio Astronomy Observatory (SARAO), a facility of the National Research Foundation, which is an agency of the Department of Science and Innovation. SARAO acknowledges the ongoing advice and calibration of GPS
systems by the National Metrology Institute of South Africa (NMISA), as well as the
time space reference systems department of the Paris Observatory. MeerTime data is housed on the OzSTAR supercomputer at Swinburne University of Technology, on which significant parts of the data reduction was performed. Parts of this research were supported by the Australian Research Council Centre of Excellence for Gravitational Wave Discovery (OzGrav), through project number CE170100004.
The Parkes radio telescope (Murriyang) is funded by the Commonwealth of Australia for operation as a National Facility managed by CSIRO. We acknowledge the Wiradjuri people as the traditional owners of the Observatory site. 
The Nan\c{c}ay Radio Observatory is operated by the Paris Observatory, associated with the French Centre National de la Recherche Scientifique (CNRS). We acknowledge financial support from the ``Programme National de Cosmologie et Galaxies'' (PNCG) and
``Programme National Hautes Energies'' (PNHE) of CNRS/INSU, France.

Parts of the data set used in this work include archived data obtained through the CSIRO Data Access Portal (http://data.csiro.au). It also made broad use of the NASA Astrophysics Data System (https://ui.adsabs.harvard.edu/).

The analysis done in this publication made use of the open source pulsar analysis packages \textsc{psrchive} \citep{PSRCHIVE}, \textsc{tempo2} \citep{tempo2} and \textsc{temponest} \citep{Lentati2014}, as well as open source \textsc{Python} libraries including Numpy, Matplotlib, Astropy and Chainconsumer.

APo and MBu acknowledge the support from the research grant "iPeska" (P.I. Andrea Possenti) funded under the INAF national call Prin-SKA/CTA approved with the Presidential Decree 70/2016. APo and MBu acknowledge that part of this work has been funded using resources from the INAF Large Grant 2022 "GCjewels" (P.I. Andrea Possenti) approved with the Presidential Decree 30/2022.
\end{acknowledgements}

\bibliographystyle{aa}
\bibliography{J1618-3921}

\end{document}